\documentclass[twocolumn]{aastex701}

\usepackage{graphicx}
\usepackage{amssymb}

\usepackage[T1]{fontenc}

\usepackage{xspace}
\newcommand{\galario}{\texttt{galario}\xspace}
\newcommand{\frank}{\texttt{frank}\xspace}

\begin{document}


\title{Probing Planet Formation within 10 au:\\Dust Substructures in Extremely Compact Protoplanetary Disks}


\correspondingauthor{Pietro Curone}

\author[orcid=0000-0003-2045-2154]{Pietro Curone}
\affiliation{Departamento de Astronomía, Universidad de Chile, Camino El Observatorio 1515, Las Condes, Santiago, Chile}
\affiliation{Instituto de Astrofísica, Pontificia Universidad Católica de Chile, Av. Vicuña Mackenna 4860, 7820436 Macul, Santiago, Chile}
\email[show]{pcurone@das.uchile.cl}  

\author[orcid=0000-0002-1078-9493]{Gijs D. Mulders} 
\affiliation{Instituto de Astrofísica, Pontificia Universidad Católica de Chile, Av. Vicuña Mackenna 4860, 7820436 Macul, Santiago, Chile}
\email[]{gijs.mulders@uc.cl} 

\author[orcid=0000-0003-2458-9756]{Nienke van der Marel} 
\affiliation{Leiden Observatory, Leiden University, PO Box 9513, 2300 RA Leiden, The Netherlands}
\email[]{nmarel@strw.leidenuniv.nl} 

\author[orcid=0000-0002-1199-9564]{Laura M. Pérez} 
\affiliation{Departamento de Astronomía, Universidad de Chile, Camino El Observatorio 1515, Las Condes, Santiago, Chile}
\email[]{lperez@das.uchile.cl}

\begin{abstract}

ALMA high-resolution surveys have so far focused mainly on large, bright protoplanetary disks, where substructures are routinely detected, leaving compact disks poorly characterized. Yet, the survival of millimeter emission in Class~II sources implies that radial drift must have been halted by pressure traps, previously undetected due to limited angular resolution. We present ALMA Band~7 (0.9~mm) continuum observations of 11 of the smallest ($R_{68} < 10$~au) Class~II disks in Lupus, reaching $0.02''$ (2~au) resolution, the highest currently available with ALMA at this wavelength. We detect substructures in 4 disks, including a clear inner cavity of radius 1.6~au, and hints of substructure in 3 more. Two disks also host faint outer rings beyond their bright inner core. Combining these results with demographic information on Class II disks observed with ALMA, we examine how the substructure detection rate varies with disk size, restricting to disks whose diameter is resolved by at least 3 resolution elements. We find that nearly all disks ($\gtrsim$95\%) larger than 30~au show substructures, compared to a still substantial $\sim$50\% between 3 and 30~au, with the lower rate likely affected by optical depth. The planet masses required to open gaps within 5~au fall in the super-Earth to mini-Neptune regime. These results demonstrate that substructures are ubiquitous across the disk population, and that planet formation conditions can be probed down to a few au from the host star.

\end{abstract}

\keywords{\uat{Protoplanetary disks}{1300} --- \uat{Dust continuum emission}{412} --- \uat{Planet formation}{1241} --- \uat{Radio interferometry medium}{1346} }


\section{Introduction} \label{sec:introduction}

High spatial resolution observations of protoplanetary disks at (sub)millimeter wavelengths have revealed that a large fraction of disks contain substructure in the form of rings and gaps \citep[e.g.,][]{Andrews2020_review}. These structures are indicative of pressure bumps in the disk gas, which halt the inward radial drift of pebbles toward the star (e.g., \citealt{Pinilla2012}). Because the timescales for unimpeded radial drift are much shorter than the protoplanetary disk lifetime \citep{Birnstiel2010}, such traps largely determine the location, amount, and lifetime of the solid mass available for planet formation. They also mark sites of active planet-building, either because an embedded planet has already created the pressure bump, as in the cases of PDS~70 \citep{Keppler2018,Haffert2019, Benisty2021} and WISPIT~2 \citep{vanCapelleveen2025,Close2025, Lawlor2026},  or because the concentration of solids at pressure maxima promotes their subsequent growth into planetesimals and protoplanets \citep[e.g.,][]{Guilera2020}.

While pressure bumps likely play a key role in disk evolution, they are not routinely detected at the scales where most planets are thought to form. The census of disk substructure is most complete for disks larger than $\sim 30$\,au \citep{vanderMarel_Mulders2021, Bosschaart2026}, yet most giant exoplanets are found between 1 and 10\,au \citep{Fulton2021}, and smaller planets likely within an au \citep{Mulders2021}.
Therefore, the structures associated with forming the majority of observed exoplanets likely reside at separations much smaller than what most ALMA surveys have probed, unless such planets formed at larger radii and later migrated inward. High-resolution ALMA programs have predominantly targeted large, bright disks \citep{Andrews18_DSHARP, Long2018, Cieza2019, Teague2025}, resolving substructures at scales of 5--30\,au at distances of 100--200\,pc. At smaller scales, substructures have been resolved down to $\sim$1\,au in TW~Hya \citep{Andrews2016}, aided by being the closest protoplanetary disk ($\sim$60\,pc). For compact disks, this regime remains largely unexplored.

Compact disks may appear small because radial drift has initially driven dust inwards in the absence of giant planets at large orbital radii. Left completely unimpeded, however, radial drift would deplete millimeter grains well before the Class~II stage \citep[e.g.][]{Pinilla2025}, yet we observe millimeter emission in these sources, suggesting the presence of close-in pressure traps blocking the drift. Notably, the planet mass required to open a detectable dust gap decreases steeply with orbital radius \citep{Sinclair2020}, so such close-in traps could be carved by planets substantially less massive than those inferred at tens of au. This is consistent with rocky planets and super-Earths being expected to form efficiently at sub-au radii in drift-dominated disks \citep{Sanchez2024}. The identification of such substructure, however, depends critically on the survey angular resolution and sensitivity \citep{Bae_2023_PPVII}. Compact disks targeted with sufficient resolution do reveal substructures \citep[e.g.,][]{Kurtovic2021, Shi2024, Guerra-Alvarado2025}, though whether pressure traps exist down to the innermost few au remains unanswered. In compact disks, such traps would reside at small radii, making them the natural laboratory to probe planet formation conditions at the orbital separations where most exoplanets are found \citep{ZhuDong2021_exoplanet_statistics}.

A near-complete, high-resolution view of the Lupus disk population has recently been provided by \citet{Guerra-Alvarado2025}, who presented ALMA Band~6 (1.3\,mm) continuum observations at $0.03\arcsec$--$0.04\arcsec$ resolution for a sample of 73 Class~II protoplanetary disks. Their survey reveals that over 67\% of Lupus disks have dust radii smaller than 30\,au. The detection rate of substructure is strongly size-dependent: while approximately 79\% of disks with $R_{\rm dust} > 30$\,au show clear rings or cavities, only $\sim$39\% of compact disks in the 8--30\,au range and $\sim$13\% of the smallest disks ($<8$\,au) do so. They report 11 new disks with substructures, including one of the smallest cavities measured to date, with a radius of 4.1\,au in Sz~72. Crucially, the majority of the smallest disks remain only marginally resolved even at $0.04\arcsec$, meaning that substructure in these sources cannot yet be ruled out.

A complementary perspective on the same compact disk population in Lupus has been offered by \citet{Ricciardi2026}, who presented ALMA Band~7 observations of 17 Lupus disks targeting $^{12}$CO and $^{13}$CO ($J=3-2$) at $0.25\arcsec$ resolution, roughly one order of magnitude deeper in sensitivity than the earlier Lupus survey by \citet{Ansdell2018}. The derived gas disk sizes are compact ($R_{\rm CO} \lesssim 40$\,au), and the gas-to-dust size ratios indicate that these disks are not drift-dominated. A similarly low gas-to-dust size ratio was independently found for the two best-resolved disks in the AGE-PRO Lupus sample \citep{Deng2025}. These results support the picture in which a substantial fraction of the faint Lupus disks are intrinsically compact rather than merely depleted of CO.

In this work, we present higher-resolution and higher-sensitivity ALMA Band~7 observations of the 11 smallest disks in the sample of \citet{Guerra-Alvarado2025}, obtained with the highest angular resolution currently available from ALMA in the C-10 configuration. In Sect.~\ref{sec:observations}, we describe the observations and data reduction. In Sect.~\ref{sec:vis_fit}, we introduce the visibility modeling. In Sect.~\ref{sec:results}, we present the substructures identified and the dust disk extent. In Sect.~\ref{sec:discussion_subs} and Sect.~\ref{sec:discussion_origin}, we discuss our results in the context of the broader disk population and the physical origin of the detected substructures, respectively. We summarize our findings in Sect.~\ref{sec:conclusions}.

\begin{figure*}
\centering
\includegraphics[width=1\hsize]{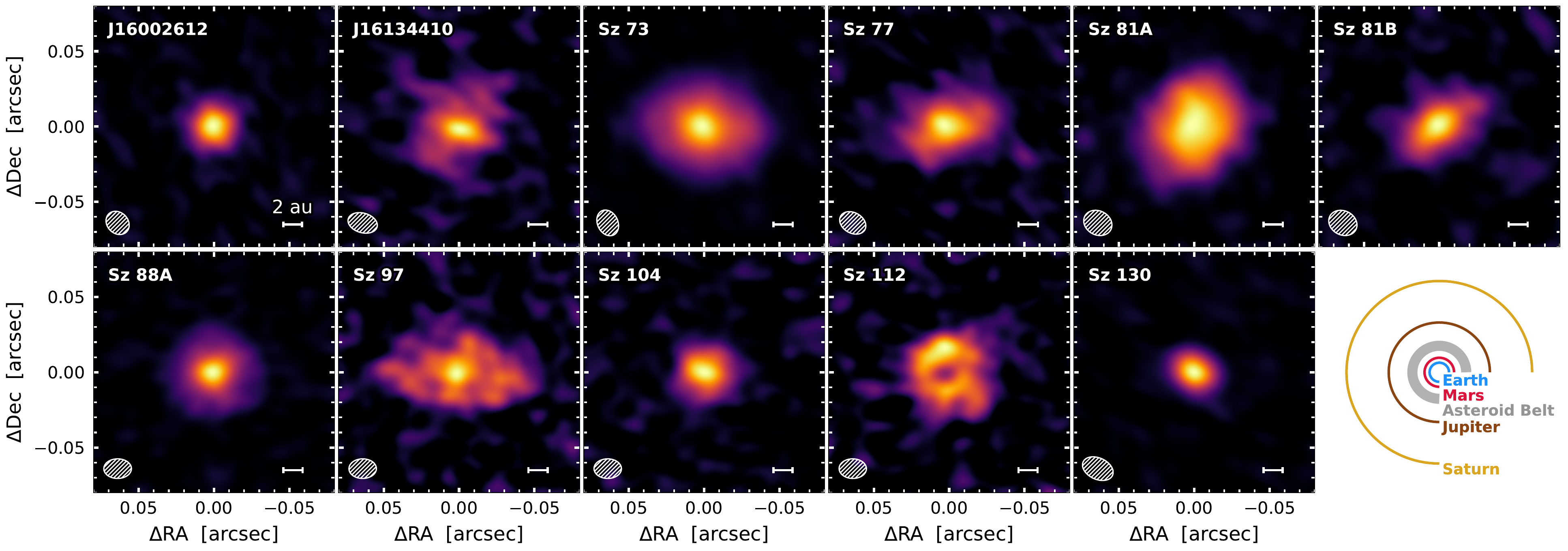}
\caption{Gallery of continuum images at 0.87 mm (343.5 GHz) of our sample, obtained with robust of 0.5. The source order is alphabetical. All images are shown on the same angular scales. The FWHM of the synthesized beams and the 2~au scale bars are indicated in the lower left and right corners of each panel, respectively. The color scale is linear and ranges from null to peak intensity for each disk. The corresponding color scales for each source are shown in Figures~\ref{fig:single_sources_subs}, \ref{fig:single_sources_hint_subs}, and \ref{fig:single_sources_no_subs}. For comparison, in the lower right are shown the angular sizes of the orbits of Earth, Mars, Asteroid Belt, Jupiter, and Saturn, assuming the average distance of the sources in our sample (157.4~pc).}
\label{fig:gallery_rob0.5}
\end{figure*}

\section{Observations and Data Reduction} \label{sec:observations}

Our sample was selected from the Class~II Lupus survey of \citet{Guerra-Alvarado2025}, from which we identified the 11 smallest disks that were marginally resolved at $0.04\arcsec$ resolution, excluding the three disks that remained unresolved even at that scale (J15592523-4235066, J16075475-3915446, J16084940-3905393). The sample consists of nine M-type and two late K-type (K7) stars. The stellar parameters and the naming convention we use for our targets are reported in Appendix~\ref{sect:appendix_supplementary}.

Observations were carried out as part of ALMA Cycle~12 program 2024.1.00583.S (PI: G.~D.~Mulders) in July and August 2025, using an average of 44 antennas in the C-10 configuration with a maximum baseline of 15.2\,km (one execution block reaching 12.8\,km). The shortest baselines ranged from 89 to 233\,m, corresponding to maximum recoverable scales of $0.23$--$0.35\arcsec$, well above the disk angular sizes of $<0.08\arcsec$ estimated from \citet{Guerra-Alvarado2025}. The spectral setup consisted of four spectral windows centered at 336.5, 338.5, 348.5, and 350.5\,GHz, each with a bandwidth of 1.875\,GHz in Time Division Mode, giving a total continuum bandwidth of 7.5\,GHz. We observed in Band~7 rather than the Band~6 used by \citet{Guerra-Alvarado2025} in order to take advantage of the increased dust emission flux density and the higher angular resolution achievable at shorter wavelengths. For eight of the ten pointings, covering nine targets since Sz~81A and Sz~81B fall within the same primary beam, we allocated 15--16 minutes of on-source integration time. The two sources with the lowest integrated fluxes in \citet{Guerra-Alvarado2025} received longer on-source integrations: 1\,hour for J16134410 and 2\,hours for J16002612. Precipitable water vapor (PWV) ranged from 0.4 to 0.9\,mm with an average of 0.6\,mm across all execution blocks. The full observing log, including phase and bandpass/flux calibrators, is reported in Table~\ref{tab:obs_log}.

\begin{deluxetable*}{lcccccc}
\tabletypesize{\footnotesize}
\tablewidth{1\textwidth} 
\tablecaption{Continuum Image Properties}
\tablehead{
\colhead{Source} & \colhead{$\theta_\mathrm{b}$} & \colhead{PA${_\mathrm{b}}$} &\colhead{rms Noise} & \colhead{Peak $I_\nu$, $T_\mathrm{b}$} & \colhead{$F_\nu$} & \colhead{$M_\mathrm{d}$}\\
\colhead{} & \colhead{(mas,$\;\;$ au)} & \colhead{(deg)} & \colhead{($\mu$Jy beam$^{-1}$,$\;\;$ K)} & \colhead{(mJy beam$^{-1}$,$\;\;$ K)} & \colhead{(mJy)} & \colhead{($M_\oplus$)}} 
\colnumbers
\startdata 
J16002612 & $17\times14, \;\; 2.7\times2.3$ & 44 & 15$,\;\;$ 5.1 & 0.41$,\;\;$ 25.2 & $1.51\pm0.13$ & $0.10$ \\
J16134410 & $20\times13, \;\; 3.1\times2.0$ & 74 & 20$,\;\;$ 5.4 & 0.30$,\;\;$ 19.0 & $1.83\pm0.20$ & $0.20$ \\
Sz 73      & $18\times14, \;\; 2.8\times2.2$ & 26 & 36$,\;\;$ 6.7 & 1.81$,\;\;$ 82.8 & $21.43\pm1.49$ & $1.84$ \\
Sz 77      & $19\times14, \;\; 2.9\times2.2$ & 60 & 38$,\;\;$ 6.6 & 0.64$,\;\;$ 32.5 & $3.95\pm0.28$ & $0.21$ \\
Sz 81A     & $20\times16, \;\; 3.1\times2.5$ & 62 & 41$,\;\;$ 6.3 & 1.02$,\;\;$ 40.7 & $8.72\pm0.35$ & $0.67$ \\
Sz 81B     & $20\times16, \;\; 3.2\times2.5$ & 62 & 41$,\;\;$ 6.3 & 0.69$,\;\;$ 29.8 & $3.72\pm0.64$ & $0.32$ \\
Sz 88A     & $19\times13, \;\; 3.0\times2.0$ & 90 & 47$,\;\;$ 7.3 & 1.82$,\;\;$ 84.3 & $8.06\pm0.44$ & $0.38$ \\
Sz 97      & $19\times13, \;\; 3.0\times2.0$ & 90 & 47$,\;\;$ 7.4 & 0.54$,\;\;$ 30.1 & $5.01\pm0.53$ & $0.52$ \\
Sz 104     & $19\times13, \;\; 3.0\times2.1$ & 90 & 47$,\;\;$ 7.3 & 0.82$,\;\;$ 42.1 & $3.02\pm0.32$ & $0.23$ \\
Sz 112     & $19\times13, \;\; 3.0\times2.1$ & 89 & 47$,\;\;$ 7.3 & 0.54$,\;\;$ 30.1 & $3.78\pm0.28$ & $0.32$ \\
Sz 130     & $22\times14, \;\; 3.5\times2.2$ & 65 & 35$,\;\;$ 6.1 & 1.29$,\;\;$ 51.2 & $5.33\pm0.48$ & $0.42$ \\
\enddata
\tablecomments{Column (1): target name. Column (2): synthesized beam FWHM major and minor axes. Column (3): synthesized beam PA. Column (4): image rms noise. Column (5): image peak intensity. Note that the noise and peak brightness temperature were computed using the full Planck law. Properties in columns (2)-(5) refer to CLEAN images with robust 0.5. Column (6): integrated flux density from robust 2.0 images with statistical uncertainty, excluding the 10\% absolute flux calibration error. Column (7): estimated lower limit for the dust mass from flux density in Column (6), assuming optically thin emission (Appendix~\ref{sec:appendix_dust_mass}).}
\label{tab:image_properties}
\end{deluxetable*}

The data reduction and calibration are described in Appendix~\ref{sec:appendix_calibration}. Continuum images were produced using the \texttt{tclean} task in CASA \citep{CASATeam2022} with Briggs weighting, adopting a multiscale deconvolver with scale sizes of $[0, 1, 2]$ synthesized beams, a cell size of $0.001\arcsec$, an image size of 2048 pixels, and cleaning down to $1\sigma$ via the \texttt{nsigma} parameter. For each source, we generated images at robust parameter values between $-2.0$ and $2.0$ in steps of 0.5 to explore the trade-off between angular resolution and sensitivity. Robust~0.5 provides the best compromise for identifying substructures in the CLEAN images, while robust~2.0 maximizes sensitivity and was used for measuring integrated flux densities and inspecting faint outer disk features (Section~\ref{subsec:outer_rings}). The gallery of our sources imaged with robust~0.5 is shown in Fig.~\ref{fig:gallery_rob0.5}. The corresponding robust~2.0 images are presented in Appendix~\ref{sect:appendix_supplementary}, which also shows the Sz~81 system, including both disks in the same field of view and their relative separation. Calibrated measurement sets and CLEAN FITS images for both weightings are publicly available (see Data Availability).

Table~\ref{tab:image_properties} reports the properties of the continuum images for each target. The rms noise was measured in an emission-free annulus centered on each disk. Integrated flux densities were measured within an elliptical mask sharing the center, position angle (PA), and aspect ratio of the disk, with a semimajor axis equal to 1.5 times the outer radius of detected emission, defined as where the $2\sigma$ contour intersects the disk major axis. For Sz~73 and Sz~130, which host faint outer rings (Sect.~\ref{subsec:outer_rings}), the masks  include those outer features. The statistical uncertainty was estimated as the standard deviation of the integrated flux measured in $\sim$10 non-overlapping elliptical masks of identical shape placed in emission-free regions of the field of view. Since all targets fall well within the primary beam, images without primary beam correction were used to ensure uniform noise. The uncertainties reported are statistical only and exclude the 10\% absolute flux calibration uncertainty in ALMA Band~7.\footnote{See Sect.~10.2.6 of the ALMA Technical Handbook: \url{https://almascience.nrao.edu/proposing/technical-handbook/}}. We also report in Table~\ref{tab:image_properties} a lower limit on the total dust mass derived from the integrated flux density as explained in Appendix~\ref{sec:appendix_dust_mass}.

\section{Visibility Fitting} \label{sec:vis_fit}

To characterize the morphologies of the observed disks, we model the visibilities following the pipeline presented in \citet{Curone2025}\footnote{The pipeline is available on GitHub: \url{https://github.com/pcurone/exoALMA_continuum_pipeline}}. This pipeline consists of two steps: a parametric fit with \galario \citep{Tazzari2018_galario} to obtain the disk geometrical parameters, which are then used as inputs to a nonparametric fit with \texttt{frankenstein} (hereafter \frank, \citealt{Jennings2020_frank}).

\subsection{\galario fit} \label{subsec:galario_fit}

\galario computes synthetic visibilities from a parametric brightness profile (1D in our case) by Fourier transform, comparing them directly to the observed visibilities at the same $uv$-points \citep{Tazzari2018_galario}. The best-fit model is determined by minimizing the $\chi^2$ via MCMC using the \texttt{emcee} sampler \citep{ForemanMackey2013_emcee}. Our goal here was not an exhaustive characterization of the disk substructures, which is instead the purpose of the \frank fit described in Section~\ref{subsec:frank_fit}, but rather to obtain robust estimates of the geometric parameters of each disk: inclination, position angle (PA), and the offsets in R.A. and decl. between the disk center and the phase center ($\Delta$RA and $\Delta$Dec).

Eight out of eleven targets (J16002612, J16134410, Sz~77, Sz~81A, Sz~81B, Sz~88A, Sz~97, Sz~104) were fitted with a single central Gaussian,
\begin{equation}
    I(R) = f_0 \,\exp{\left[-\frac{R^2}{2\sigma_0^2}\right]}.
\end{equation}
For Sz~73 and Sz~130, the presence of extended outer emission required adding a Gaussian ring component,
\begin{equation}
    I(R) = f_0 \,\exp{\left[-\frac{R^2}{2\sigma_0^2}\right]} + f_1 \,\exp{\left[-\frac{(R-r_1)^2}{2\sigma_1^2}\right]}.
\end{equation}
For Sz~112, which shows a central cavity, we used a single Gaussian ring without a central Gaussian component. All parameters were sampled with uniform priors; the flux normalizations $f_0$ and $f_1$ were sampled uniformly in logarithmic space. Each run used 100 walkers and converged within 5000 iterations. The best-fit results are reported in Appendix~\ref{sect:appendix_supplementary}.

\subsection{\frank fit} \label{subsec:frank_fit}

To characterize the intensity radial profiles, we used \frank \citep{Jennings2020_frank}. It reconstructs the 1D axisymmetric intensity radial profile by modeling it as a Fourier-Bessel series and using a discrete Hankel transform to compute synthetic visibilities, which are then fitted directly to the deprojected, centered, unbinned observed visibilities within a Bayesian framework employing a Gaussian process for regularization. The fit is nonparametric and operates at sub-beam resolution, recovering features that would remain undetected in the CLEAN image or its azimuthally averaged profile, while exploiting the full sensitivity of the data \citep{Jennings2022_DSHARP}.

We performed all fits in logarithmic intensity space (\texttt{method=LogNormal}), which enforces non-negativity and suppresses high-frequency oscillations compared to fitting in linear space. The hyperparameter $R_{\rm max}$, beyond which \frank assumes zero emission, was fixed at $0.5\arcsec$, well beyond the extent of all sources in the sample. The number of radial grid points was set to $N = 500$, and the power spectrum normalization to $p_0 = 10^{-35}$, the standard value for logarithmic fits. For each disk, we explored a grid of values for the SNR threshold parameter $\alpha \in [1.05, 1.10, 1.15, 1.20, 1.30]$ and the smoothing parameter $w_{\rm smooth} \in [10^{-4}, 10^{-3}, 10^{-2}, 10^{-1}]$. The combination $\alpha = 1.15$ and $w_{\rm smooth} = 10^{-3}$ was found to provide the best balance between avoiding overfitting and preserving small-scale features, and was adopted uniformly for all targets.

The geometric parameters (inc, PA, $\Delta$RA, $\Delta$Dec) used to deproject the visibilities were taken from the \galario posteriors (Sect.~\ref{subsec:galario_fit}). Uncertainties on the \frank profiles and on all derived radii (Section~\ref{subsec:radii}) were obtained via Monte Carlo error propagation, similarly to \citet{Carvalho2024} and \citet{Curone2025}: we ran \frank 500 times, each time drawing the geometric parameters from a randomly selected iteration in the last 1000 steps of the \galario MCMC chain. The resulting 16th--84th percentile range reflects the statistical uncertainty propagated from the geometric parameter posteriors, and does not include systematic uncertainties associated with the choice of \frank hyperparameters.

\section{Results} \label{sec:results}

\begin{figure*}
\centering
\includegraphics[width=1\hsize]{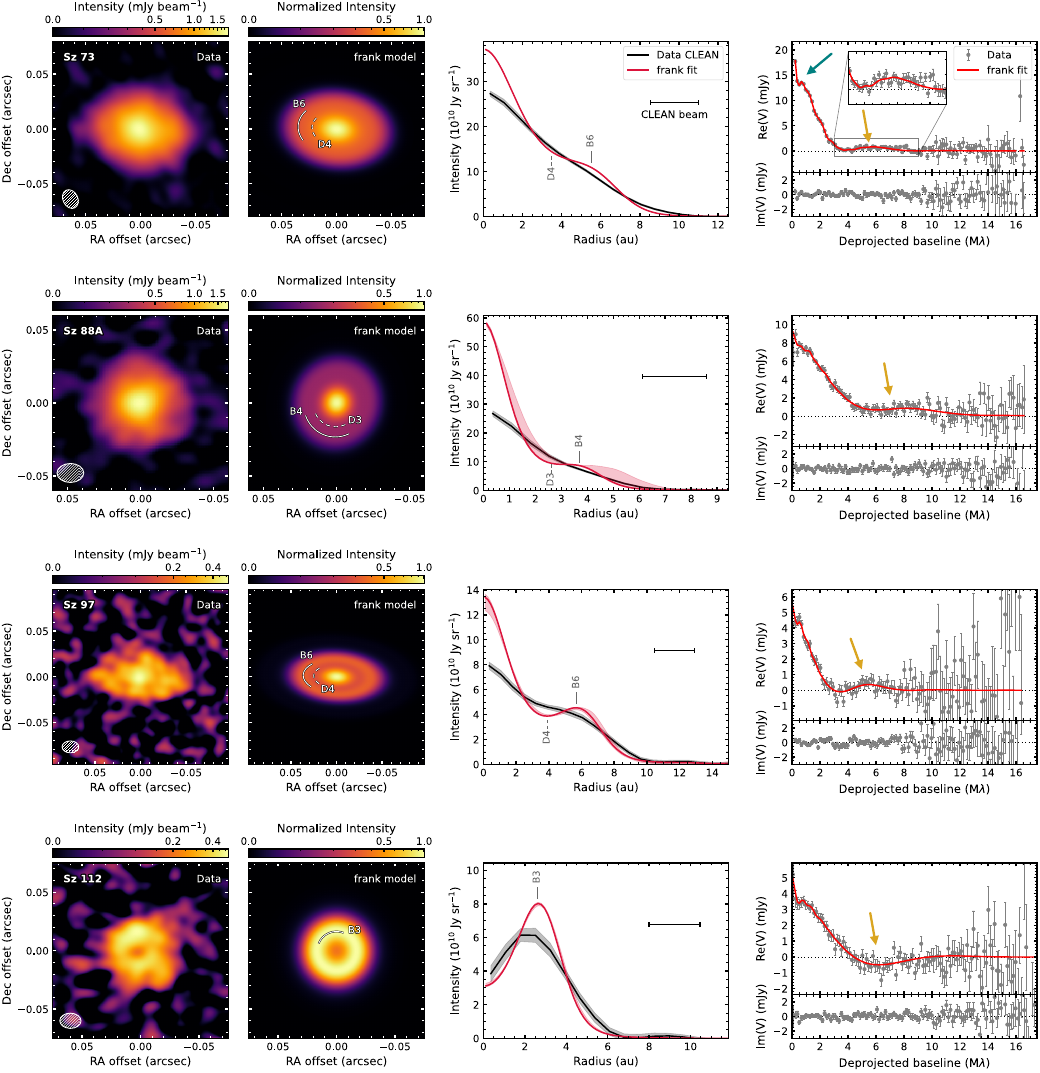}
\caption{Comparison of data and \frank\ visibility fit for disks showing substructures within 10~au. \textit{First panel:} Continuum image (robust $0.5$) with the synthesized beam shown in the lower left; an asinh stretch is applied to the color scale to visually enhance the fainter emission. \textit{Second panel:} \frank\ model swept over $2\pi$ and projected, with normalized intensity and asinh stretch; dashed arcs labeled "D" mark gaps and solid arcs labeled "B" mark rings, each followed by the distance from the star in au. \textit{Third panel:} Deprojected, azimuthally averaged radial intensity profiles of the CLEAN data with robust~0.5 (black) and \frank\ model (red). Gray shading shows the $1\sigma$ CLEAN uncertainty per radial bin divided by the square root of the number of beams in the annulus; red shading shows the 16th--84th percentile range of the Monte Carlo uncertainty propagation of the \frank\ fit (Sect.~\ref{subsec:frank_fit}). Rings and gaps are marked as in the second panel; the scale bar indicates the mean FWHM of the synthesized beam. \textit{Fourth panel:} Deprojected visibilities (real and imaginary parts) azimuthally averaged in 150~k$\lambda$ bins for the data (gray points) and \frank\ model (red line). Arrows indicate robust detections (solid) or hints (dashed) of inner (yellow) and outer (teal) disk substructures.}
\label{fig:single_sources_subs}
\end{figure*}

\begin{figure*}
\centering
\includegraphics[width=1\hsize]{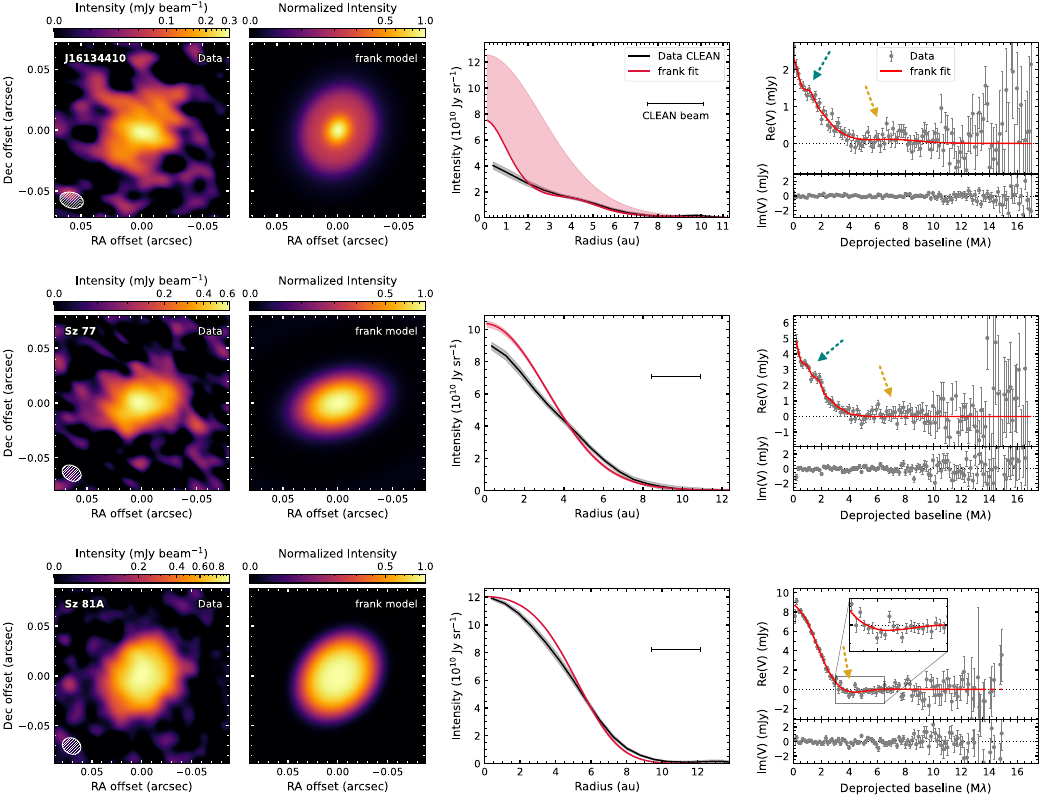}
\caption{Same as Fig.~\ref{fig:single_sources_subs}, but for sources with hints of substructures within 10 au.}
\label{fig:single_sources_hint_subs}
\end{figure*}

\begin{figure*}
\centering
\includegraphics[width=1\hsize]{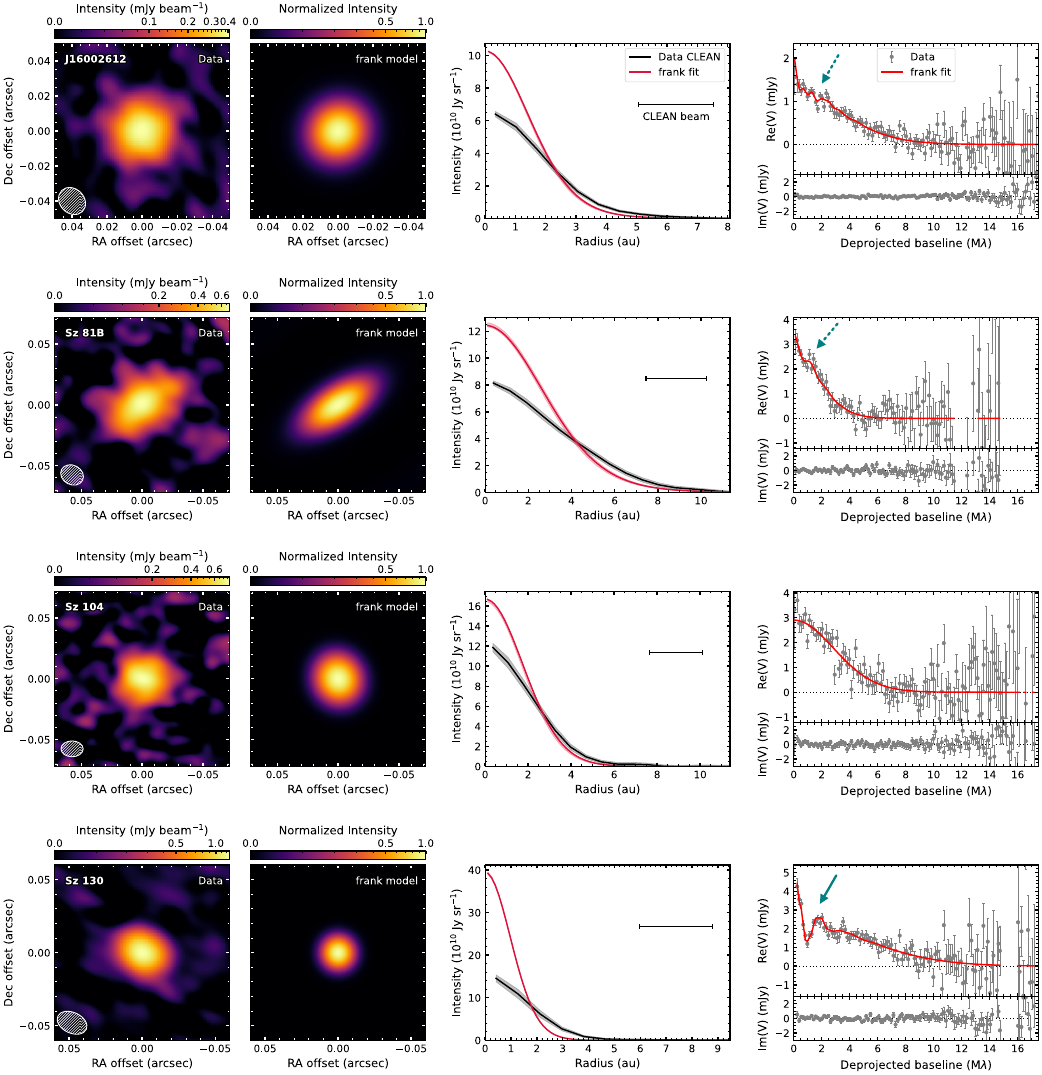}
\caption{Same as Fig.~\ref{fig:single_sources_subs}, but for sources with no substructures appearing within 10 au.}
\label{fig:single_sources_no_subs}
\end{figure*}

\subsection{Inner Substructures}
\label{subsec:inner_subs}

We present the substructure content of the compact disks in our sample. We use the term "inner substructures" to refer to features appearing within 10~au, distinguishing them from the faint outer rings discussed in Sect.~\ref{subsec:outer_rings}. We divide our sample into three groups: disks with robust detections of substructures, disks with hints of substructures, and smooth disks with no detected substructures. These three groups are presented in Figures~\ref{fig:single_sources_subs}, \ref{fig:single_sources_hint_subs}, and \ref{fig:single_sources_no_subs}, respectively. Each figure shows the CLEAN image alongside the \frank model swept over $2\pi$ and projected, their intensity radial profiles, and the deprojected visibilities (real and imaginary parts) plotted as a function of deprojected baseline length. The intensity radial profile of the CLEAN image was obtained with the tool \texttt{GoFish} \citep{Teague2019}.
Given that we are detecting substructures on scales of a few au in extremely compact and faint disks, our classification of substructure fidelity is based primarily on the visibility data: we assess whether coherent departures from a smooth profile are present at deprojected baselines in the range of 4--10~M$\lambda$, and whether these features are consistently reproduced by the \frank fit.

\textit{Disks with substructures.} This group includes four disks with robust substructures within 10~au; their properties are reported in Table~\ref{tab:substructures}. The clearest case is Sz~112: already in the image plane, the presence of an inner cavity is evident. A cavity in Sz~112 had been predicted from SED modeling \citep{vanderMarel2016}, but ALMA observations at $\sim$30~mas ($\sim$5~au) resolution did not resolve it \citep{vanderMarel2022}. At our $\sim$2~au resolution, we resolve the cavity both in the image plane and in the \frank model: the inner edge of the ring, defining also the radial extent of the cavity,  is located at 1.6~au, with the ring peak at 2.6~au. Sz~97 also shows a substructure visible in the image plane and confirmed by both the visibilities and the \frank model: beyond a centrally peaked emission component, a gap-ring pair is present at radii of 3.9 and 5.7~au, respectively. For Sz~112 and Sz~97, we identify annular substructures in the \frank best-fit intensity radial profiles following the procedures explained in Appendix~\ref{sec:appendix_subs_prop}. Rings are labeled "B" (bright) and gaps "D" (dark).

Despite showing no evident substructures in the image plane, Sz~73 and Sz~88A exhibit clear substructures within 10~au in the visibility data and their \frank models. Unlike Sz~112 and Sz~97, where the features correspond to well-defined local minima and maxima of the intensity profile, in these two sources the profiles show plateau-like features, possibly because optical depth effects suppress the contrast of the underlying surface density substructure (see Sect.~\ref{subsec:optical_depth}). The criteria to identify plateaus and derive their properties are presented in Appendix~\ref{sec:appendix_subs_prop}.

\textit{Disks with hints of substructures.} This group includes sources where clear substructures are not detected, but where features in the visibility data or the image plane prevent a classification as purely smooth. J16134410 is the second faintest source in the sample in terms of total flux. Beyond the centrally peaked emission, it shows low-SNR extended emission in the image plane. In the visibility domain, the real part approaches zero at $\sim$4~M$\lambda$, rises again between 6 and 8~M$\lambda$, and then declines toward zero between 8 and 10~M$\lambda$ before the noise becomes dominant beyond $\sim$10~M$\lambda$. This behavior is interpreted by the best-fit \frank model as a plateau-like feature at 3--4~au. However, the Monte Carlo error propagation shows that varying the geometric parameters within the 16th--84th percentile range of the \galario posterior distribution substantially changes the resulting \frank profile (red shaded area in the \frank intensity radial profile in Fig.~\ref{fig:single_sources_hint_subs}), so we caution against a definitive interpretation of this feature in J16134410. Sz~77 shows two hints of substructure. First, the real part of the visibilities in the 5--10~M$\lambda$ baseline range is on average slightly above zero, a feature not reproduced by the \frank model or by any realization in the Monte Carlo error propagation, which consistently produce a smooth profile. Second, the image shows a possible point-like feature on the southeast side of the disk, at a significance of $4\sigma$, which also appears in the residuals obtained by subtracting the \frank model from the data (Fig.~\ref{fig:gallery_resid}). This feature is broadly co-spatial with a non-axisymmetric shoulder visible in the eastern side of the lower-resolution Band~6 image from \citet{Guerra-Alvarado2025}; a direct comparison is shown in Fig.~\ref{fig:Sz77_comparison}. Sz~81A appears smooth in the CLEAN image but shows visibilities that clearly fall below zero at $\sim$4~M$\lambda$, a behavior reproduced by the best-fit \frank model. Negative real visibilities at these baselines can arise from the presence of a central cavity (as for \ Sz~112), and the \frank intensity profile indeed shows a notably flat central emission component, possibly indicative of a shallow and/or unresolved inner cavity. 

\textit{Disks with no substructures.} This group comprises the sources where no substructure within 10~au is detected, either in the image plane or in the visibility data: J16002612, Sz~81B, Sz~104, and the inner emission component of Sz~130. Three of these four sources (Sz~130, J16002612, and Sz~104) are the most compact disks in our sample, all with $R_{68} < 3$~au (Sect.~\ref{subsec:radii}), making it inherently more difficult to resolve substructure at our $\sim$2~au resolution. Sz~81B, while slightly larger ($R_{68} = 4.5$~au), has the highest inclination in the sample at $61.7\degr$, which reduces the effective resolution along the minor axis, limiting the detectability of axisymmetric features in the $uv$-plane.

\subsection{Faint Outer Rings}
\label{subsec:outer_rings}

While our program primarily targets substructures within $\sim$10~au, two sources in the sample, Sz~73 and Sz~130, show faint outer rings beyond that radius. We note, however, that our C-10 configuration observations have a maximum recoverable scale (MRS) of $0.23-0.35\arcsec$ ($36-55$~au) in Band~7, comparable to the radial extent of the features discussed below. The properties of these outer rings should therefore be interpreted with caution, since spatial filtering can affect extended emission at these scales.

\begin{figure}
\centering
\includegraphics[width=1\columnwidth]{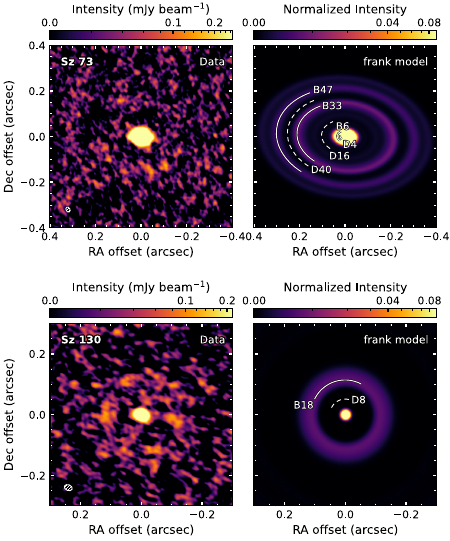}
\caption{Faint outer rings in our sample. Left: saturated robust 2.0 image of the data. Right: saturated \frank\ model swept over $2\pi$ and projected, with normalized intensity; dashed arcs labeled "D" mark gaps and solid arcs labeled "B" mark rings, each followed by the distance from the star in au. Both have an asinh stretch.}
\label{fig:outer_rings}
\end{figure}

Figure~\ref{fig:outer_rings} shows the robust 2.0 images, saturated to highlight these faint features, alongside their \frank models. The outer rings are also evident as coherent features in the real part of the visibility profiles at baselines below $\sim$2~M$\lambda$, indicated by solid teal arrows in Figs.~\ref{fig:single_sources_subs} and \ref{fig:single_sources_no_subs} for Sz~73 and Sz~130, respectively. The \frank models recover two gap-ring pairs in Sz~73, with rings at 33 and 47~au, and one gap-ring pair in Sz~130, with a ring at 18~au. The properties of these outer substructures are reported in Table~\ref{tab:substructures}. A faint outer ring at $\sim$40~au with a width of $\sim$10~au in Sz~73 was already detected in the lower-resolution Band~6 data of \citet{Guerra-Alvarado2025}. Their single ring is spatially coincident with our two rings at 33 and 47~au, consistent with the spatial filtering caveat noted above.

\begin{figure*}
\centering
\includegraphics[width=1\hsize]{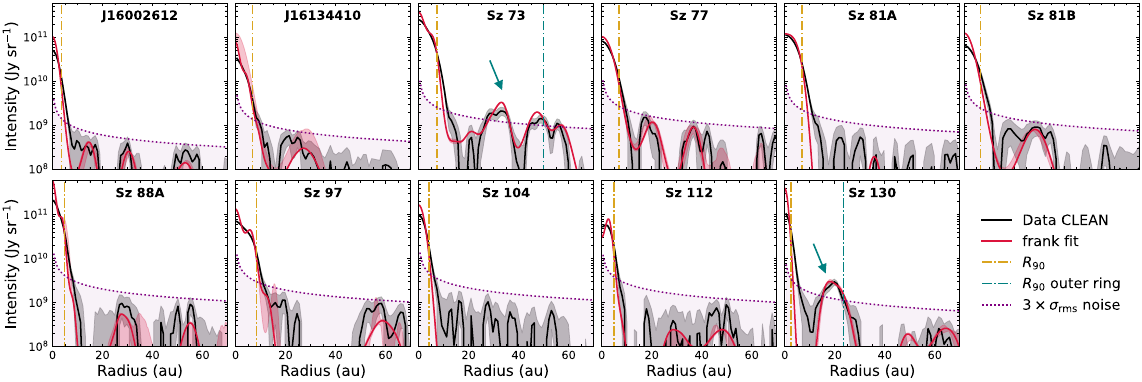}
\caption{Gallery of azimuthally averaged radial intensity profiles on a log-linear scale for each source. The deprojected profile from robust 2.0 CLEAN images (black solid line) and the \frank model (red solid line) are shown, with uncertainties as described in Fig.~\ref{fig:single_sources_subs}. Dash-dotted gold vertical lines mark $R_{90}$ for each disk. For Sz~73 and Sz~130, dash-dotted teal vertical lines mark $R_{90}$ computed including their faint outer rings, indicated by teal arrows. The purple shaded region marks intensity levels below $3\sigma_{\rm rms}$, where $\sigma_{\rm rms}$ is the rms noise of the CLEAN image scaled by the square root of the number of beams within each radial annulus to account for the radial dependence of the azimuthal average.}
\label{fig:int_profiles_outer_rings}
\end{figure*}

Figure~\ref{fig:int_profiles_outer_rings} shows the azimuthally averaged intensity radial profiles of the CLEAN data and the \frank model on a log-linear scale, extending to 70~au. The outer rings in Sz~73 and Sz~130 are the only outer features detected above $3\sigma$ in both the CLEAN image and the \frank model, where $\sigma$ is scaled to account for the radial dependence of the azimuthal average. Hints of additional outer features are present in the intensity profiles and in the short-baseline visibility data of four other sources, namely J16002612, J16134410, Sz~77, Sz~81B, (dashed teal arrows in Figs.~\ref{fig:single_sources_hint_subs} and \ref{fig:single_sources_no_subs}; profiles in Fig.~\ref{fig:int_profiles_outer_rings}), but deeper observations with more complete short-baseline coverage are needed to assess their nature.

The flux density of the outer ring emission, integrated from the \frank profile over radii of 15--65~au for Sz~73 and 8--40~au for Sz~130, is 8.6 and 3.0~mJy, respectively, corresponding to 40\% and 56\% of the total flux density of each disk (Table~\ref{tab:image_properties}). Given the large radii and low surface brightness of these features, the assumption of optically thin emission is well justified, and applying Eq.~\ref{eq:dust_mass}  computed as described in Appendix~\ref{sec:appendix_dust_mass} yields dust masses of $0.74$~$M_\oplus$ for Sz~73 and $0.24$~$M_\oplus$ for Sz~130.

\subsection{Dust Radii} \label{subsec:radii}

All disks in our sample are spatially resolved. The real part of the deprojected visibilities in Figs.~\ref{fig:single_sources_subs}, \ref{fig:single_sources_hint_subs}, and \ref{fig:single_sources_no_subs} consistently starts at positive values and decreases toward zero at longer baselines. The declining trend confirms that all disks are spatially resolved, while the convergence toward zero, rather than a positive floor, rules out any significant unresolved central component.

We compute dust radii from the \frank intensity profiles, which provide an unconvolved, nonparametric reconstruction of the disk emission, free from the dependence on imaging choices such as the Briggs weighting  that affects CLEAN-based profiles. The outermost radius is determined by a "first plateau" of the cumulative flux profile: starting from the disk center and moving outward, we identify the first radius at which the derivative of the cumulative flux falls to 5\% of its maximum value. We adopt this criterion rather than the more common first-peak method because, for some sources, the cumulative does not exhibit a well-defined local maximum even when a plateau is clearly present; additionally, this approach is directly applicable to the \frank profile, which is strictly monotonic by construction. The radii obtained in this way represent the extent of the inner bright emission component. For Sz~73 and Sz~130, which host faint outer rings (Sect.~\ref{subsec:outer_rings}), we additionally report radii computed including those outer regions, where the maximum radius is set by the global maximum of the cumulative flux. The resulting radii are listed in Table~\ref{tab:disk_radii}.



\begin{deluxetable*}{lccc}
\tabletypesize{\footnotesize}
\tablewidth{1\textwidth}
\tablecaption{Dust Disk Radii \label{tab:disk_radii}}
\tablehead{
\colhead{Source} & \colhead{$R_{68}$} & \colhead{$R_{90}$} & \colhead{$R_{95}$} \\
\colhead{} & \colhead{(au, mas)} & \colhead{(au, mas)} & \colhead{(au, mas)}
}
\colnumbers
\startdata
J16002612 & $2.50^{+0.06}_{-0.03}, \;\; 15.5^{+0.4}_{-0.2}$ & $3.60^{+0.07}_{-0.06}, \;\; 22.3^{+0.4}_{-0.4}$ & $4.13^{+0.06}_{-0.08}, \;\; 25.6^{+0.4}_{-0.5}$ \\
J16134410 & $4.54^{+0.28}_{-0.12}, \;\; 28.9^{+1.8}_{-0.7}$ & $6.57^{+0.31}_{-0.16}, \;\; 41.8^{+2.0}_{-1.0}$ & $7.57^{+0.55}_{-0.19}, \;\; 48.2^{+3.5}_{-1.2}$ \\
Sz 73     & $5.71^{+0.02}_{-0.02}, \;\; 36.5^{+0.1}_{-0.1}$ & $7.27^{+0.02}_{-0.02}, \;\; 46.4^{+0.1}_{-0.1}$ & $7.90^{+0.02}_{-0.02}, \;\; 50.5^{+0.2}_{-0.2}$ \\
Sz 73 incl. outer ring    & $28.6^{+0.2}_{-0.2}, \;\; 183^{+1}_{-1}$ & $49.9^{+0.2}_{-0.2}, \;\; 319^{+1}_{-1}$ & $55.7^{+0.3}_{-0.3}, \;\; 356^{+2}_{-2}$ \\
Sz 77     & $4.84^{+0.09}_{-0.08}, \;\; 31.2^{+0.6}_{-0.5}$ & $6.86^{+0.17}_{-0.14}, \;\; 44.3^{+1.1}_{-0.9}$ & $7.83^{+0.23}_{-0.18}, \;\; 50.6^{+1.5}_{-1.2}$ \\
Sz 81A    & $5.29^{+0.04}_{-0.04}, \;\; 33.7^{+0.3}_{-0.3}$ & $6.92^{+0.06}_{-0.05}, \;\; 44.1^{+0.4}_{-0.3}$ & $7.60^{+0.07}_{-0.05}, \;\; 48.4^{+0.5}_{-0.3}$ \\
Sz 81B    & $4.50^{+0.08}_{-0.09}, \;\; 28.6^{+0.5}_{-0.6}$ & $6.53^{+0.11}_{-0.12}, \;\; 41.4^{+0.7}_{-0.7}$ & $7.51^{+0.12}_{-0.13}, \;\; 47.6^{+0.8}_{-0.8}$ \\
Sz 88A    & $3.05^{+0.72}_{-0.04}, \;\; 19.4^{+4.6}_{-0.3}$ & $3.93^{+0.91}_{-0.06}, \;\; 25.0^{+5.8}_{-0.4}$ & $4.31^{+0.90}_{-0.07}, \;\; 27.4^{+5.7}_{-0.4}$ \\
Sz 97     & $6.57^{+0.10}_{-0.09}, \;\; 42.1^{+0.6}_{-0.6}$ & $8.39^{+0.11}_{-0.51}, \;\; 53.7^{+0.7}_{-3.3}$ & $9.77^{+0.18}_{-1.35}, \;\; 62.6^{+1.2}_{-8.7}$ \\
Sz 104    & $2.82^{+0.07}_{-0.04}, \;\; 17.9^{+0.4}_{-0.2}$ & $3.97^{+0.10}_{-0.06}, \;\; 25.1^{+0.6}_{-0.4}$ & $4.51^{+0.12}_{-0.06}, \;\; 28.5^{+0.7}_{-0.4}$ \\
Sz 112    & $3.77^{+0.06}_{-0.03}, \;\; 23.9^{+0.4}_{-0.2}$ & $4.75^{+0.07}_{-0.03}, \;\; 30.1^{+0.4}_{-0.2}$ & $5.25^{+0.06}_{-0.04}, \;\; 33.3^{+0.4}_{-0.3}$ \\
Sz 130    & $1.68^{+0.02}_{-0.01}, \;\; 10.7^{+0.1}_{-0.1}$ & $2.35^{+0.03}_{-0.01}, \;\; 14.9^{+0.2}_{-0.1}$ & $2.65^{+0.05}_{-0.01}, \;\; 16.8^{+0.3}_{-0.1}$ \\
Sz 130 incl. outer ring    & $19.47^{+0.23}_{-0.09}, \;\; 123.3^{+1.5}_{-0.6}$ & $23.50^{+0.36}_{-0.11}, \;\; 148.8^{+2.3}_{-0.7}$ & $25.01^{+0.49}_{-0.22}, \;\; 158.4^{+3.1}_{-1.4}$ \\
\enddata
\tablecomments{Column (1): target name. Columns (2)--(4): dust disk radii enclosing 68\%, 90\%, and 95\% of the total flux, respectively, derived from the \frank intensity profile (see Sect.~\ref{subsec:radii}). Uncertainties represent the 16th--84th percentile range from Monte Carlo sampling of the galario geometric parameter posteriors.}
\end{deluxetable*}

\section{Substructure Demographics and Dust Evolution} \label{sec:discussion_subs}

\subsection{Substructure Detection Rate} \label{subsec:subs_occ_rate}

To place our results in the broader context of disk substructure demographics, we compiled an updated catalog of continuum substructure detections from all major ALMA observations and surveys of protoplanetary disks published up to June 2026, restricted to Class~II sources (publicly available; see Data Availability). Starting from the sample of \citet{Bae_2023_PPVII}, we added results from \citet{vanderMarel2022}, \citet{Yamaguchi2024}, \citet{Huang2024}, \citet{Shi2024}, \citet{Guerra-Alvarado2025}, AGE-PRO \citep{Vioque2025}, exoALMA \citep{Curone2025}, ODISEA \citep{Dasgupta2025, Shoshi2025, Bhowmik2026}, \citet{Bosschaart2026}, individual sources \citep{Ribas2025, Facchini2026}, images from the ALMA archive, and this work. The catalog contains information from different ALMA Bands, and for each source records the measured $R_{68}$, the angular resolution of the observations, and whether at least one substructure was detected. We classify as substructured any disk showing (by eye) at least one clear feature beyond a purely smooth profile, including rings, gaps, cavities, and asymmetries, as well as softer features such as shoulders; for sources from the literature, we adopt the classification reported in the corresponding paper. For the eleven disks in our sample, only those with clear substructure detections (Sz~73, Sz~88A, Sz~97, and Sz~112; Fig.~\ref{fig:single_sources_subs}) are classified as substructured, while the three disks with hints of substructures (Fig.~\ref{fig:single_sources_hint_subs}) are conservatively included in the unstructured category, along with the four smooth disks (Fig.~\ref{fig:single_sources_no_subs}). Their $R_{68}$ values are taken from Table~\ref{tab:disk_radii}. For Sz~73 and Sz~130, we adopt the radii computed including their faint outer rings. 

A crucial diagnostic to consider when inspecting the substructure detection rate on a population level is the effective angular resolution introduced by \citet{Bae_2023_PPVII}, $\theta_D/\theta_{\rm res}$, where $\theta_D = 2R_{68}$ is the disk diameter and $\theta_{\rm res}$ the FWHM of the angular resolution of the observation. This quantity captures how well-resolved each disk is relative to the observing beam, enabling a fair comparison of substructure detection rates across surveys conducted at different angular resolutions. A comparison between our updated catalog and the results of \citet{Bae_2023_PPVII} is presented in Appendix~\ref{sec:comparison_bae2023}, showing that our updated catalog has a higher substructure detection rate at low ($\theta_D/\theta_{\rm res} < 3$) and intermediate ($3 \leq \theta_D/\theta_{\rm res} \leq 10$) effective resolutions.

\begin{figure}
\centering
\includegraphics[width=1\hsize]{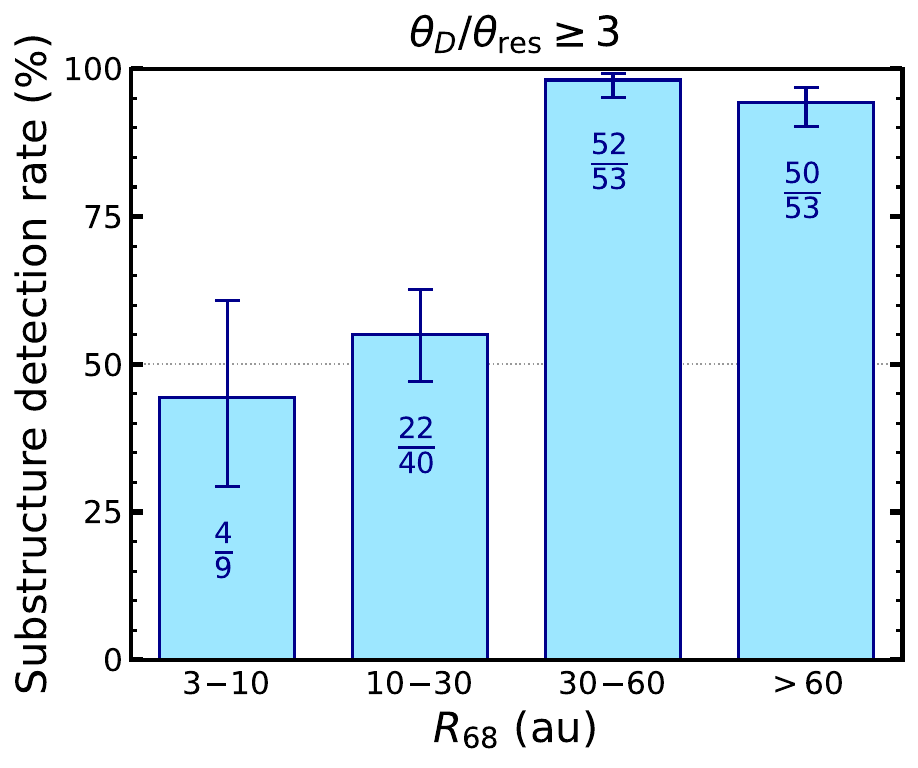}
\caption{Substructure detection rate as a function of $R_{68}$ for sources with effective angular resolution $\theta_D/\theta_{\rm res} \geq 3$. Bar heights show the observed fraction of disks with detected substructure, with the label  giving the number of substructured disks over the total number of disks in that bin. Uncertainties represent $1\sigma$ Wilson score confidence intervals.}
\label{fig:subs_occ_rate_vs_R68}
\end{figure}

To study how the substructure detection rate varies with disk size, we focus on sources with $\theta_D/\theta_{\rm res} \geq 3$ in our updated catalog, the minimum effective resolution required to reliably detect substructures \citep{Bae_2023_PPVII}. We divide the sample into four size bins: extremely compact ($R_{68}<10$~au), compact ($10 \leq R_{68} < 30$~au), intermediate ($30 \leq R_{68} < 60$~au), and large ($R_{68} \geq 60$~au).
The result is shown in Fig.~\ref{fig:subs_occ_rate_vs_R68}.

Intermediate and large disks ($R_{68} > 30$~au) show substructure detection rates of 98\% and 94\%, respectively. Notably, the rate does not drop substantially for smaller disks: the two smallest bins, 3--10~au and 10--30~au (the former mainly populated by our survey), retain detection rates of 44\% and 55\%, respectively, well above the near-zero rates found in low effective resolution surveys. This supports the picture in which the apparent scarcity of substructures in compact disks is primarily an observational bias driven by insufficient angular resolution, rather than a genuine physical absence. The lower rates in the two smallest bins compared to the larger ones may partly reflect optical depth effects, which suppress substructure contrast in the dense inner regions of compact disks (Sect.~\ref{subsec:optical_depth}).

\subsection{Substructures and Trapping}
\label{subsec:subs_trapping}

In the absence of pressure bumps, pebbles drift inward on timescales much shorter than the typical disk lifetime of a few Myr \citep{Birnstiel2010}, raising the question of why Class~II compact disks still retain detectable millimeter emission. The substructures we detect in 4 out of 11 disks, and tentatively in 3 more, are consistent with pressure traps present even at orbital radii of a few au, capable of halting the inward drift of pebbles and retaining solid mass over the disk lifetime \citep{Pinilla2012}.

This finding is consistent with the significant substructure detection rates found even in the smallest disk size bins at matched effective angular resolution (Sect.~\ref{subsec:subs_occ_rate}). Rather than being a population of drift-dominated, trap-free disks, compact disks appear to host the same types of pressure traps as their larger counterparts, with the difference that these traps are located closer to the host star. This is in line with the scenario proposed by \citet{Stadler2022}, in which the outermost pressure trap sets the disk size. Compact disks would then simply be those whose outermost trap resides at a few au, rather than at tens of au as in larger disks.

For the smooth disks in our sample where no substructure is detected, several interpretations remain open. Substructures could be present but unresolved at our $\sim$2~au resolution, or at contrasts below our sensitivity limit. Alternatively, optical depth effects may suppress the observable contrast of underlying surface density features in the dense inner regions of these compact disks (Sect.~\ref{subsec:optical_depth}). Finally, some of these sources may be genuinely drift-dominated, gradually depleting their solid mass reservoir without efficient trapping (see Sect.~\ref{subsec:radial_drift}).

\subsection{Dust Radial Drift}
\label{subsec:radial_drift}

A direct test of the strength of radial drift in our sample is offered by the gas-to-dust size ratio, $R_{\rm CO}/R_{\rm dust}$. Observed values across several star-forming regions cluster around 2.5--3.0 \citep{Sanchis2021, Long2022, Pinilla2025}, while theoretical and numerical models predict that ratios above $\sim$4 indicate strong radial drift \citep{Trapman2019, Toci2021}. Five disks in our sample overlap with \citet{Ricciardi2026}, who measured $R_{68}$ from $^{12}$CO $J=3-2$ line emission, allowing a direct comparison with our dust radii. In the following, we denote the dust radius as $R_{68,\rm dust}$, identical to the $R_{68}$ reported throughout the rest of this work (Table~\ref{tab:disk_radii}), and the CO-derived radius from \citet{Ricciardi2026} as $R_{68,\rm CO}$. Before this comparison, we first verify that our $R_{68,\rm dust}$ measurements are robust and not significantly affected by the lack of short baselines in our dataset, which could resolve out faint extended emission.

\begin{figure*}
\centering
\includegraphics[width=1\hsize]{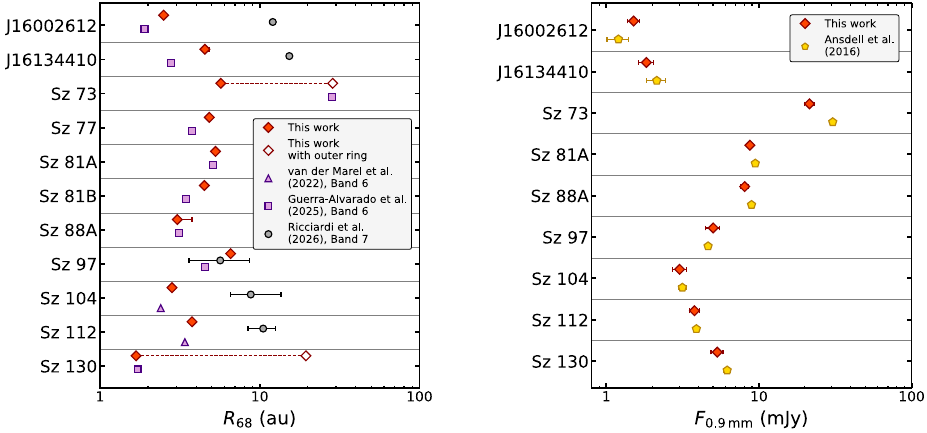}
\caption{Comparison between estimates from this work and literature values for dust $R_{68}$ (left) and flux density (right) at ALMA Band 7 (0.9~mm).}
\label{fig:radii_flux_comparison}
\end{figure*}

The left panel of Fig.~\ref{fig:radii_flux_comparison} compares our $R_{68,\rm dust}$ with literature estimates: Band 6 values from \citet{vanderMarel2022} and \citet{Guerra-Alvarado2025} (${\sim}0.03\arcsec$ resolution), and Band 7 values from \citet{Ricciardi2026} (${\sim}0.25\arcsec$ resolution). All literature radii come from \galario\ Gaussian fits, while ours are derived from \frank models. We find good agreement with the high-resolution Band 6 estimates, with our values occasionally slightly larger. This could either reflect a wavelength effect, with Band~6 probing slightly larger grains than Band~7, which are more strongly affected by radial drift and therefore more centrally concentrated, or, more likely, be a consequence of the \galario Gaussian fits adopted in the literature, which are dominated by the bright central emission and can miss faint, non-Gaussian outer flux that the non-parametric \frank\ fit recovers. Interesting are the two disks with faint outer rings: for Sz~73, our radius including the outer ring matches \citet{Guerra-Alvarado2025}, who also detect and fit the outer emission, while for Sz~130 only the inner-disk radius agrees, since \citet{Guerra-Alvarado2025} do not recover the outer ring we detect.

A larger discrepancy emerges in the comparison with \citet{Ricciardi2026}: among the five overlapping disks, only Sz~97 is consistent, while the other four (J16002612, J16134410, Sz~104, Sz~112) have radii larger by a factor of 3--5. Two explanations are possible: either \citet{Ricciardi2026} lack the resolution to fully resolve the dust emission, or their shorter baselines recover faint extended flux that our longer-baseline data filter out. To test the second possibility, the right panel of Fig.~\ref{fig:radii_flux_comparison} compares our integrated flux densities with other Band~7 (0.9~mm) measurements from the literature. \citet{Ansdell2016} report flux densities for 9 of our 11 sources, measured in the image plane at ${\sim}0.3\arcsec$ resolution, consistent with our estimates for all but Sz~73, for which we measure $21.43\pm1.49$~mJy against their $30.43\pm0.55$~mJy. This $\sim$9~mJy difference likely reflects the spatial filtering of the extended outer ring emission of Sz~73, given the MRS limitation of our C-10-only baseline coverage discussed in Sect.~\ref{subsec:outer_rings}. Significant loss of large-scale flux appears unlikely for the rest of the sample, and the radius discrepancy with \citet{Ricciardi2026} is more plausibly due to their lower angular resolution.

Having established the reliability of our $R_{68,\rm dust}$ measurements, we compare them with $R_{68,\rm CO}$ from \citet{Ricciardi2026} in Table~\ref{tab:rgas_rdust}. For J16134410, only an upper limit on $R_{68,\rm CO}$ is available, giving $R_{68,\rm CO}/R_{68,\rm dust} < 3.6$, consistent with weak radial drift. Sz~104 has an upper limit of $R_{68,\rm CO}/R_{68,\rm dust} < 4.9$, which does not provide a meaningful constraint. J16002612 and Sz~112 show high ratios of $13.5 \pm 6.1$ and $7.5 \pm 2.4$, respectively, but \citet{Ricciardi2026} flag both as affected by cloud contamination so these values should be treated with caution. Sz~97 is the only overlap source free of this caveat and, notably, one of the four disks in our sample with a clear dust substructure (Sect.~\ref{subsec:inner_subs}). Its ratio, $1.9 \pm 1.0$, indicates that radial drift is not the dominant mechanism setting $R_{68,\rm dust}$, exactly as expected if the detected gap-ring pair has locally halted the drift and kept dust and gas co-spatial, even in a disk as small as $R_{68,\rm dust} = 6.57$~au. An additional caveat is that even fainter, undetected outer rings could still be present beyond the ones already identified in Sz~73 and Sz~130, which would push $R_{68,\rm dust}$ to higher values and lower these ratios.

\begin{deluxetable}{lccc}
\tabletypesize{\footnotesize}
\tablewidth{1\textwidth}
\tablecaption{Gas-to-Dust Disk Size Ratio of overlapping targets between \cite{Ricciardi2026} and this work.}  \label{tab:rgas_rdust}
\tablehead{
\colhead{Source} & \colhead{$R_{68,\rm dust}$} & \colhead{$R_{68,\rm CO}$} & \colhead{$R_{68,\rm CO}/R_{68,\rm dust}$} \\
\colhead{} & \colhead{(au)} & \colhead{(au)} & \colhead{}
}
\colnumbers
\startdata
J16002612 & $2.50^{+0.06}_{-0.03}$ & $33.7^{+15.2}_{-15.2}$ & $13.5^{+6.1}_{-6.1}$ \\
J16134410 & $4.54^{+0.28}_{-0.12}$ & $<16.5$ & $<3.6$ \\
Sz 97     & $6.57^{+0.10}_{-0.09}$ & $12.7^{+6.7}_{-6.7}$  & $1.9^{+1.0}_{-1.0}$ \\
Sz 104    & $2.82^{+0.07}_{-0.04}$ & $<13.9$ & $<4.9$ \\
Sz 112    & $3.77^{+0.06}_{-0.03}$ & $28.3^{+9.2}_{-9.2}$  & $7.5^{+2.4}_{-2.4}$ \\
\enddata
\tablecomments{Column (1): target name. Column (2): dust $R_{68}$ (this work, see Sect.~\ref{subsec:radii}). Column (3): $R_{68}$ of the integrated $^{12}$CO line flux, from \cite{Ricciardi2026}. Column (4): ratio of gas to dust radii.}
\end{deluxetable}

\subsection{Optical Depth}
\label{subsec:optical_depth}

Estimating the optical depth of the dust emission in our sources would require multiwavelength observations to measure the spectral index (e.g., \citealt{Carrasco-Gonzalez2019}), which are not available for this sample. \citet{Guerra-Alvarado2025} provide a quantitative estimate of the mean optical depth, $\bar\tau$, for the Lupus disk population from a grid of radiative transfer models at Band 6 (1.3~mm). For five of our targets they report $\bar\tau = 0.27$, $1.03$, $0.72$, $1.56$, and $5.00$ for Sz~73, Sz~77, Sz~81B, Sz~97, and J16134410, respectively, already pointing to marginally or fully optically thick emission for most of these sources. Since dust opacity increases toward shorter wavelengths, with the values adopted in this work and in \citet{Guerra-Alvarado2025} (3.5 and 2.3~cm$^2$~g$^{-1}$ at 0.87 and 1.3~mm, respectively, both from \citealt{Beckwith1990}) implying a factor of $\sim$1.5 increase, the optical depth of our Band~7 observations is expected to be even higher than these Band~6 estimates.

This is consistent with the expectation that the innermost regions of these compact disks, where dust is concentrated within a few au of the host star, reach surface densities high enough to make the Band~7 emission optically thick. In this regime, the observed intensity traces the dust temperature distribution rather than the surface density, naturally suppressing the contrast of any underlying substructure. This is also consistent with the plateau-like features identified in Sz~73 and Sz~88A (Sect.~\ref{subsec:inner_subs}), where the \frank profiles show no well-defined local minima and maxima but rather a flattening of the intensity slope.

For the sources where no substructure is detected, optical depth effects cannot be excluded as a contributing factor. A direct analog is offered by PDS~66, where Band~6 (1.3~mm) and Band~7 (0.9~mm) observations appear smooth while Band~3 (3.0~mm) observations reveal annular substructures \citep{Ribas2025}: at longer wavelengths the emission becomes more optically thin, revealing structure hidden at shorter wavelengths. Observations at longer wavelengths would therefore be valuable for our sample. However, since the achievable angular resolution scales linearly with wavelength, the highest angular resolution currently accessible with ALMA at Band~3 is $\sim$50~mas, corresponding to $\sim$8~au at the distance of our sources, insufficient to probe the few-au scales relevant here. Resolving the innermost disk regions at optically thinner wavelengths will require either an ALMA extension with longer baselines or the Next Generation Very Large Array (ngVLA).

\subsection{Occurrence of Faint Outer Rings} \label{subsec:discussion_outer_rings}

High-resolution observations of Taurus disks have shown that the inner emission cores of extended ring disks are  indistinguishable in surface brightness, radius, and radial profile shape from compact disks that appear smooth at the same resolution (\citealt{Long2019}, \citealt{Facchini2019}), with the larger size of ring disks set entirely by additional rings beyond the inner core. This raises the question of whether  compact disks really lost their outer dust through rapid inward drift, or they instead retain faint outer rings below current sensitivity limits. Our sample confirms the second scenario for two of the eleven targets: Sz~73 and Sz~130 both retain an outer ring (Sect.~\ref{subsec:outer_rings}). Four other sources (J16002612, J16134410, Sz~77, and Sz~81B) show similar hints of emission, falling below the $3\sigma$ threshold in the intensity profiles (Fig.~\ref{fig:int_profiles_outer_rings}) and appearing as coherent features in the short-baseline visibilities (Figs.~\ref{fig:single_sources_hint_subs}, \ref{fig:single_sources_no_subs}). Confirming these requires observations with higher sensitivity and better baseline coverage at compact baselines, to recover these larger spatial scales.

\section{Origin of the Substructures} \label{sec:discussion_origin}

\subsection{Putative Planets}
\label{subsec:planets}

One natural interpretation of the observed gaps and cavities is that they are carved by embedded planets. Assuming each substructure hosts a single planet, we ask what planet population would be required to reproduce them. A first estimate could come from the observed gap width via the Hill radius relation of \citet{Lodato2019}, $\Delta_{\rm gap}=kR_{\rm Hill}$ with $k\simeq5.5$. We do not use this relation, for two reasons. First, it is calibrated on gaps at tens to hundreds of au around stars of $0.3$-$2\,M_\odot$ \citep{Lodato2019}, a regime far from our $\lesssim 10$~au gaps around $0.1$-$0.7\,M_\odot$ stars. Second, since $M_P\propto\Delta_{\rm gap}^3$, any error on the gap width from limited angular resolution propagates into the inferred mass as its cube, and at our $\sim2$~au resolution most gap widths in Table~\ref{tab:substructures} are not safely resolved.

We instead use two mass thresholds that depend only on orbital radius $a$ and stellar mass $M_\star$, not on the gap width, and are therefore immune to this resolution bias. \citet{Sinclair2020} (see their Eq.~15) combine dusty hydrodynamical simulations with radiative transfer and synthetic ALMA observations to derive the minimum planet mass $M_{\rm min}$ that produces a detectable gap ($>10\%$ decrease in intensity) in the dust continuum,
\begin{equation}
\frac{M_{\rm min}}{M_\oplus} = \left(\frac{M_\star}{M_\odot}\right)^{-0.33}\left(\frac{a}{\rm au}\right)^{3/4},
\end{equation}
where $M_\star$ is the stellar mass and $a$ is the planet semi-major axis. We also compute the pebble isolation mass $M_{\rm iso}$, the mass at which the planet reverses the gas pressure gradient just outside its orbit, creating a local pressure maximum that halts the inward drift of pebbles (Eq.~12 in \citealt{Lambrechts2014}), with the stellar-mass dependence calibrated by Eq.~19 in \citet{Sinclair2020}:
\begin{equation}
\frac{M_{\rm iso}}{M_\oplus} = 20\left(\frac{a}{5\,{\rm au}}\right)^{3/4}\left(\frac{M_\star}{M_\odot}\right)^{-0.275}.
\end{equation}
By construction $M_{\rm min}<M_{\rm iso}$: a planet below $M_{\rm iso}$ can still carve a detectable dust gap without fully halting the pebble flux across its orbit. 

\begin{figure*}
\centering
\includegraphics[width=1\hsize]{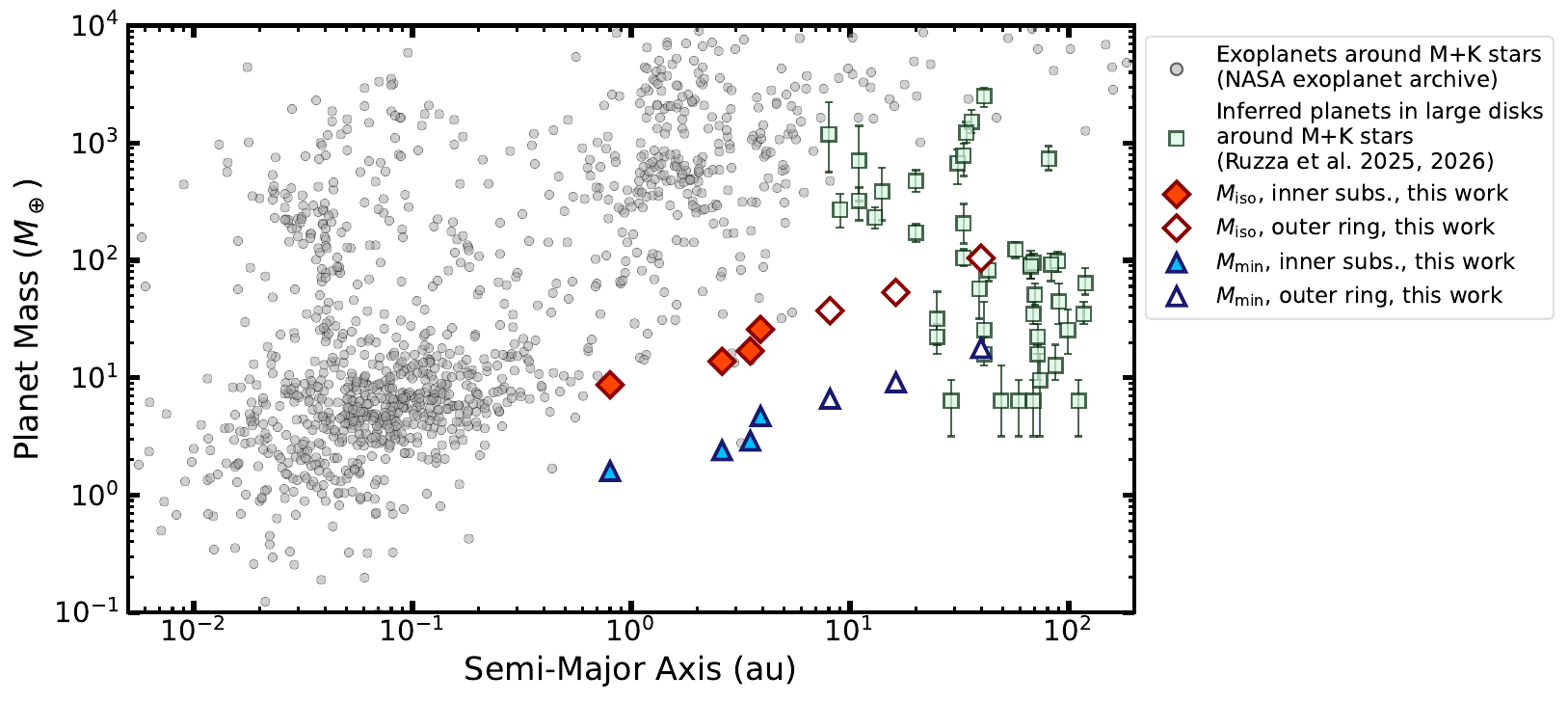}
\caption{Minimum gap-opening mass $M_{\rm min}$ \citep{Sinclair2020} and
pebble isolation mass $M_{\rm iso}$ \citep{Lambrechts2014,Sinclair2020}
for each gap and cavity in our sample (Table~\ref{tab:planetmasses}), with filled and open symbols distinguishing, respectively, inner substructures and faint outer rings (Table~\ref{tab:substructures}). Values are compared
to the population of confirmed exoplanets around M and K stars (NASA
Exoplanet Archive, \citealt{Christiansen2025_NASA_exoplanet_archive})
and to inferred planet masses in larger disks around M and K
stars \citep{Ruzza2025,Ruzza2026}.}
\label{fig:putative_planets}
\end{figure*}

Figure~\ref{fig:putative_planets} and Table~\ref{tab:planetmasses} report $M_{\rm min}$ and $M_{\rm iso}$ for every gap and cavity in our sample. Values range from $1.6$ to $18\,M_\oplus$ for $M_{\rm min}$ and from $9$ to $104\,M_\oplus$ for $M_{\rm iso}$, well in the super-Earth to Saturn-mass regime. Figure~\ref{fig:putative_planets} compares these masses to the known population of exoplanets around M and K stars (NASA Exoplanet Archive, \citealt{Christiansen2025_NASA_exoplanet_archive}) and to the planets inferred using the machine learning tool DBNets2.0 in large and bright protoplanetary disks around M and K stars \citep{Ruzza2025, Ruzza2026}. The putative planets in our sample fall between the close-in, low-mass planets found by transit and short-period radial velocity surveys and the wide-separation giants found by long-baseline radial velocity and direct imaging, a region of parameter space current surveys are not yet sensitive to \citep{Lodato2019,Bae_2023_PPVII}. The same gap is only partly filled by the proposed planets in larger disks \citep{Ruzza2025,Ruzza2026}, which probe orbital separations beyond $10$~au. The inner substructures identified here extend this within $5$~au, with $M_{\rm min}=1.6$-$4.7\,M_\oplus$ and $M_{\rm iso}=8.7$-$25.8\,M_\oplus$, positioning them as possible signatures of planet formation in the super-Earth to mini-Neptune mass regime, at orbital separations not previously reached by planet inferences based on dust substructures.

\subsection{Non-Planetary Origin of the Inner Substructures} \label{subsec:other-mechanisms}

In this Section, we discuss whether the inner substructures within 10~au that we observe can be explained by a non-planetary origin. We follow the  classification of \cite{Bae_2023_PPVII}, examining whether the hydrodynamic, magnetohydrodynamic, and dust-induced mechanisms, together with gravitational instability and stellar flybys, can account for the substructures in Table~\ref{tab:substructures}. Most of these mechanisms are disfavored or untestable with continuum-only data. The Rossby wave instability \citep{Lovelace1999}, convective overstability \citep{Klahr2014}, zombie vortex instability \citep{Marcus2015}, gravitational instability \citep{Kratter2016}, and stellar flybys \citep{Cuello2019} all predict non-axisymmetric vortices or spirals rather than the axisymmetric rings we observe. On the other hand, the vertical shear instability \citep{Flock2017}, and the Hall effect \citep{Kunz2013}, the dominant non-ideal MHD channel at these radii, both predict feature widths of order the local pressure scale height $H$, estimated to be $\approx 0.1$--$0.3$\,au for these inner substructures, narrower than the $1.0$--$2.2$\,au widths measured here. Icelines are an unlikely cause as well:  for these low-luminosity stars, the H$_2$O iceline lies within $\lesssim 1$\,au and the CO iceline beyond $\gtrsim 8$\,au \citep{Zhang2015}, not coinciding with the observed substructures at ${\sim}$2--6\,au. Overall, the substructures in Sz~73, Sz~88A, Sz~97, and Sz~112 are consistent with a planetary origin, but our single-wavelength observations, at moderate sensitivity, are not sufficient to rule out other physical mechanisms. The one structure that stands out is Sz~112, the only source with a central cavity, for which internal photoevaporation offers a natural inside-out clearing mechanism \citep{Ercolano2021}. However, population synthesis models find that photoevaporation would open an inner cavity in the gas only after most of the mm-dust has already drifted inward, unless something else is trapping it \citep{Appelgren2023}. Another possibility for the cavity is a close brown dwarf companion, given the inferred total system mass of only 0.14\,M$_\odot$. The RUWE (renormalized unit weight error) measured by Gaia is nonetheless below the 1.4 threshold for a good single-star solution \citep{Lindegren2021}, as for all sources in our sample (Table~\ref{tab:stellar_parameters}).

\section{Conclusions} \label{sec:conclusions}

We present ALMA Band~7 (0.9~mm) observations of a sample of 11 of the smallest disks ($R_{68} < 10$~au) in the Lupus star-forming region, selected from the Class~II Lupus survey of \citet{Guerra-Alvarado2025} as the smallest disks marginally resolved at $0.04\arcsec$ resolution. We model these disks through visibility fitting, using a pipeline that combines a parametric fit with \galario\ to constrain the disk geometrical parameters and a nonparametric fit with \frank\ to characterize the intensity radial profile and identify substructures. Our main findings are as follows:

\begin{itemize}
    \item We find inner substructures within 10~au in 4 disks (Sz~73, Sz~88A, Sz~97, Sz~112), hints of substructure in 3 (J16134410, Sz~77, Sz~81A), and smooth profiles in the remaining 4 (J16002612, Sz~81B, Sz~104, Sz~130). Notably, Sz~112 hosts a clear inner cavity with a radius of 1.6~au.
    
    \item Two sources (Sz~73 and Sz~130) present faint outer rings (beyond 15~au) surrounding a bright inner disk with $R_{68}$ within 6~au. Four other disks (J16002612, J16134410, Sz~77, Sz~81B) present hints of faint outer rings.

    \item All disks in our sample are spatially resolved, with $R_{68}$ within 7~au for every disk, excluding the faint outer rings in Sz~73 and Sz~130.

    \item Combining our results with demographic information on Class~II disks observed with ALMA, we study how the dust substructure detection rate varies with disk radius, controlling for the effective angular resolution $\theta_D/\theta_{\rm res}$, where $\theta_D$ is the disk diameter and $\theta_{\rm res}$ is the FWHM of the angular resolution of the observation. Considering $\theta_D/\theta_{\rm res} \geq 3$, we find substructures in nearly all intermediate (30--60~au, 98\%) and large ($>60$~au, 94\%) disks. Crucially, smaller disks still show a considerable detection rate: the compact (10--30~au) and extremely compact (3--10~au, mainly populated by our survey) bins retain a considerable detection rate of 55\% and 44\%, respectively. This indicates that the apparent scarcity of substructures in compact disks reflects insufficient angular resolution rather than a genuine physical absence, with optical depth effects possibly contributing to the lower rates in the two smallest bins.

    \item Substructures are therefore present even in the smallest disks, and their smaller sizes might simply reflect an outermost pressure trap located at only a few au from the central star.

    \item Comparing our dust radii with the CO radii measured by \citet{Ricciardi2026} for 5 of the disks in our sample, we find that 3 sources (J16134410, Sz~97, Sz~104) are not drift-dominated, while two sources show high $R_{68,\rm CO}/R_{68,\rm dust}$ ratios, $13.5\pm6.1$ for J16002612 and $7.5\pm2.4$ for Sz~112, indicative of strong radial drift.

    \item Assuming that each observed substructure is carved by a single planet, we estimate the corresponding minimum gap-opening mass $M_{\rm min}$ \citep{Sinclair2020} and pebble isolation mass $M_{\rm iso}$ \citep{Lambrechts2014,Sinclair2020}, finding values from $1.6$ to $18\,M_\oplus$ for $M_{\rm min}$ and from $9$ to $104\,M_\oplus$ for $M_{\rm iso}$, at semi-major axes of 1 to 40~au. These putative planets fill a region of parameter space that current exoplanet surveys around M and K stars are not yet sensitive to, with the inner substructures within $5$~au possibly probing the formation of super-Earths and mini-Neptunes.
\end{itemize}

These results show that pressure traps can operate down to a few au from the host star even in the smallest protoplanetary disks, with direct implications for where planet formation takes place. Observations with comparably high angular resolution from ALMA or future facilities such as the ngVLA, extended to a larger number of extremely compact disks in different star-forming regions and at other wavelengths, and complemented by information from gas tracers, will allow us to further characterize these innermost planet-forming regions.

\begin{acknowledgments}

We thank the referee whose constructive comments have improved this manuscript.
The authors are grateful to Osmar M. Guerra-Alvarado for his contribution to the ALMA proposal from which this work originates. P.C. thanks Jaehan Bae for providing the data presented in \cite{Bae_2023_PPVII}.
This paper makes use of the following ALMA data: ADS/JAO.ALMA\#2024.1.00583.S. ALMA is a partnership of ESO (representing its member states), NSF (USA) and NINS (Japan), together with NRC (Canada), MOST and ASIAA (Taiwan), and KASI (Republic of Korea), in cooperation with the Republic of Chile. The Joint ALMA Observatory is operated by ESO, AUI/NRAO and NAOJ. 

P.C., G.D.M., and L.P. acknowledge support from the ANID BASAL project FB210003. G.D.M. acknowledges support from FONDECYT project 1252141 and the "Programa de Inserción Académica 2025 de la Vicerrectoría Académica y Prorrectoría de la Pontificia Universidad Católica de Chile".  L.P. acknowledges support from ANID FONDECYT Regular 1262272.
This work made use of the Geryon cluster at the Centro de Astro-Ingenieria UC, which is supported by ANID BASAL project FB21000, BASAL CATA PFB-06, Anillo ACT-86, FONDEQUIP AIC-57, and QUIMAL 130008.
\end{acknowledgments}





\facilities{ALMA}

\software{\texttt{astropy} \citep{astropy2022},  
          \texttt{CASA} v6.6.6 \citep{CASATeam2022}, 
          \texttt{emcee} \citep{ForemanMackey2013_emcee},
          \texttt{frankenstein} \citep{Jennings2020_frank},
          \texttt{galario} \citep{Tazzari2018_galario},
          \texttt{GoFish} \citep{Teague2019},
          \texttt{matplotlib} \citep{hunter2007_matplotlib},
          \texttt{numpy} \citep{harris2020_numpy}, 
          \texttt{scipy} \citep{Virtanen2020_scipy}.
          }

\section*{Data Availability} \label{sec:data_availabity}

The data products presented in this paper, including the calibrated measurement sets, the robust 0.5 and 2.0 CLEAN FITS images, the CLEAN and frank intensity radial profiles, and the compiled catalog of protoplanetary disks with substructure classifications, dust radii, and angular resolutions of their observations, are publicly available on Zenodo at  \href{https://zenodo.org/records/20559660}{10.5281/zenodo.20559660}. The raw ALMA data are publicly available on the ALMA Science Archive \url{https://almascience.nrao.edu/aq/} under project ID 2024.1.00583.S.

\appendix
\apptablenumbers

\section{Supplementary Tables and Figures} \label{sect:appendix_supplementary}

\begin{figure*}
\centering
\includegraphics[width=1\hsize]{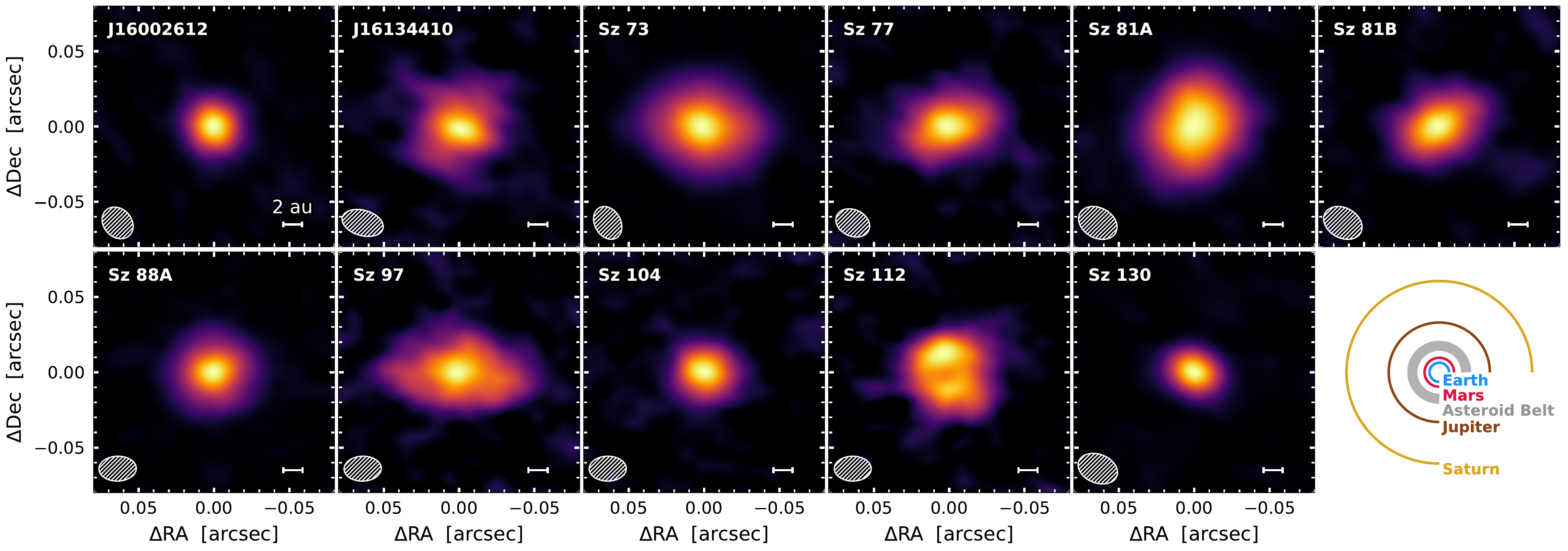}
\caption{Same as Fig.~\ref{fig:gallery_rob0.5} but for robust 2.0.}
\label{fig:gallery_rob2.0}
\end{figure*}

\begin{figure}
\centering
\includegraphics[width=0.5\hsize]{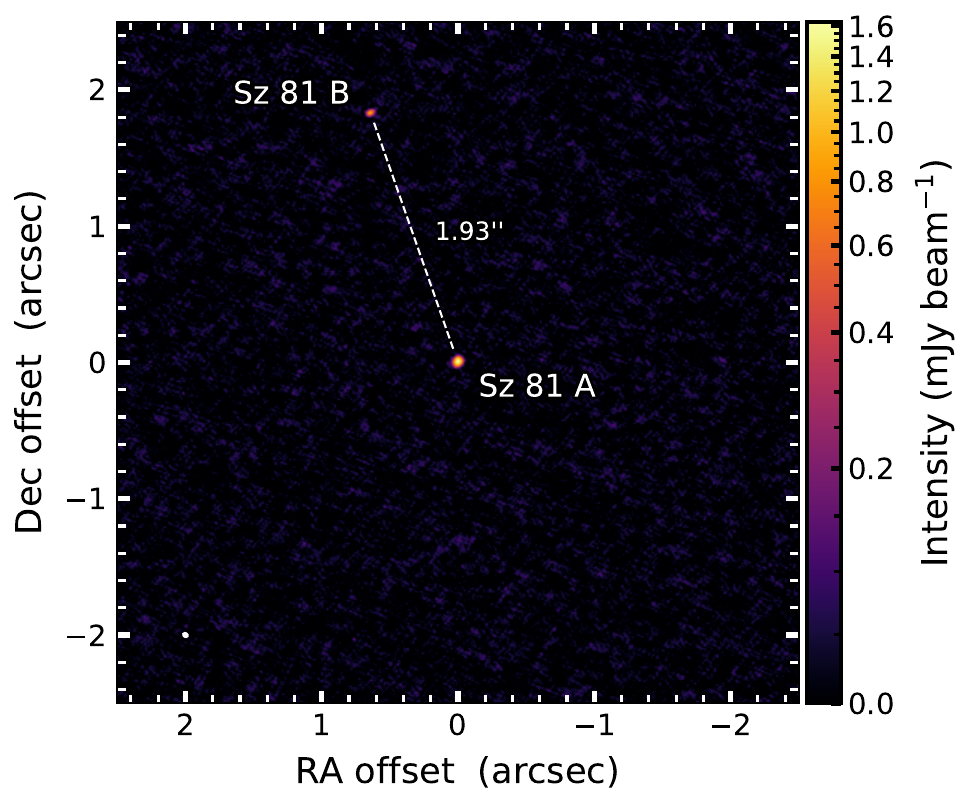}
\caption{Wide-field view of the Sz~81 system, including both A and B. The image was obtained with robust~2.0. The relative separation projected on the sky between the two components is $1.93\arcsec$, corresponding to 304~au at their average distance of 157.2~pc. The position angle of B relative to A, measured from North to East, is 19.5~deg.}
\label{fig:Sz81AB}
\end{figure}

\begin{figure*}
\centering
\includegraphics[width=1\hsize]{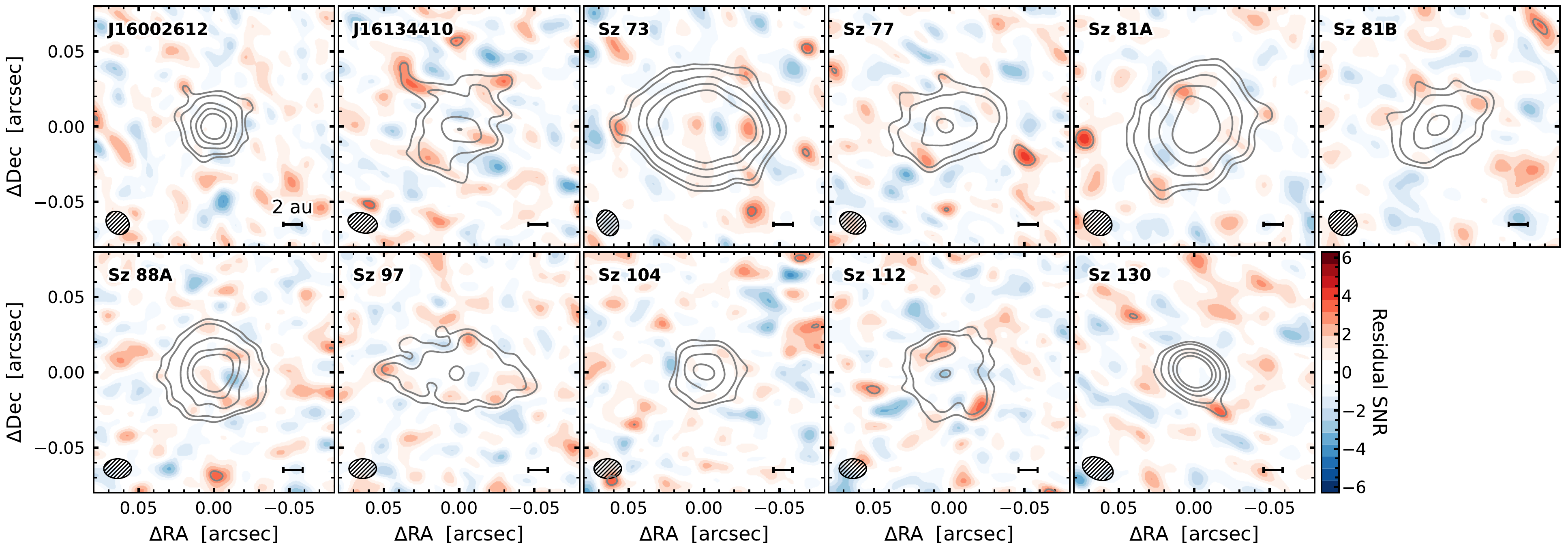}
\caption{Gallery of residuals plots generated subtracting the \frank model (sampled at the same uv-points of the observations) from the data and then imaged with CLEAN and robust 0.5. The color scale represents the values of residual SNR in units of the observed noise ($\sigma_\mathrm{rms}$) for the respective observation, and the same extremes are applied to each panel. Gray contours show the  [3, 5, 10,15, 20]$\sigma_\mathrm{rms}$ levels of the data images obtained with robust 0.5 and shown in Fig.~\ref{fig:gallery_rob0.5}. The FWHM of the synthesized beams and the 2 au scale bars are indicated in the lower left and right corners of each panel, respectively. }
\label{fig:gallery_resid}
\end{figure*}

\begin{figure}
\centering
\includegraphics[width=0.7\hsize]{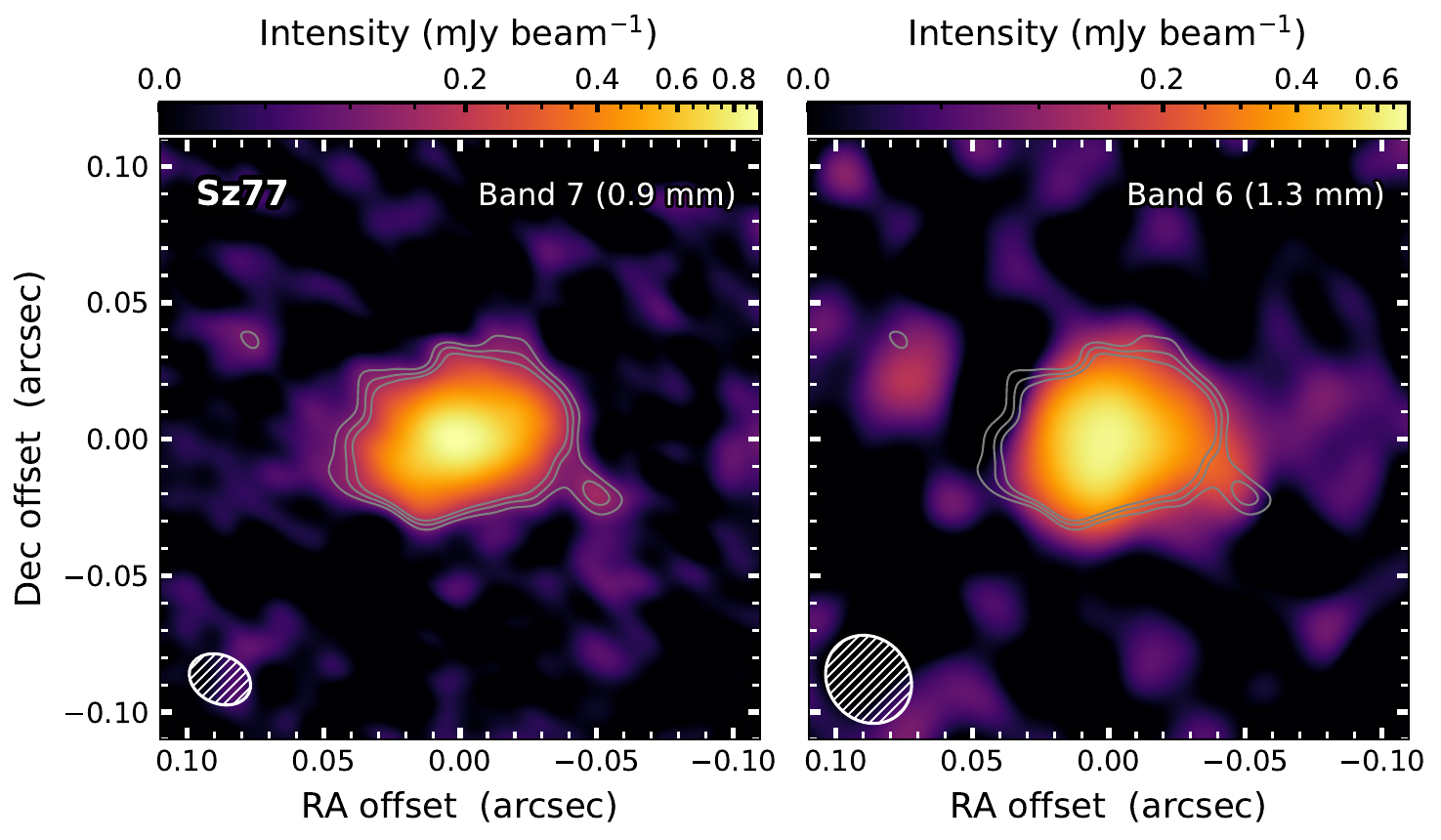}
\caption{Comparison between the robust~2.0 CLEAN image from the Band~7 data presented in this work (left) and the Band~6 CLEAN image from \citet{Guerra-Alvarado2025} (right) for Sz~77. Contours represent the $[3, 4, 5]\,\sigma_{\rm rms}$ levels of the left image. The non-axisymmetric feature reaches a $4\sigma$ significance level in our data and is co-located with a non-axisymmetric feature visible in the \citet{Guerra-Alvarado2025} image.}
\label{fig:Sz77_comparison}
\end{figure}

Here we present tables and figures that complement the findings reported in the main text. Table~\ref{tab:stellar_parameters} lists the stellar parameters of our sample. Table~\ref{tab:obs_log} summarizes the details of the ALMA observations. Figure~\ref{fig:gallery_rob2.0} shows the robust 2.0 version of the gallery in Figure~\ref{fig:gallery_rob0.5}. Figure~\ref{fig:Sz81AB} shows the Sz~81 system, including both A and B in the field of view, and their relative separation and position angle. Table~\ref{tab:galario_results} reports the results of the \galario fits (see Sect.~\ref{subsec:galario_fit}). Figure~\ref{fig:gallery_resid}
presents the gallery of \frank model residuals. Figure~\ref{fig:Sz77_comparison} compares our image of Sz~77 with the one presented in \cite{Guerra-Alvarado2025}. Table~\ref{tab:planetmasses} reports the
putative planet masses for each identified substructure (Sect.~\ref{subsec:planets}).

\begin{deluxetable*}{lccccccccc}
\tabletypesize{\footnotesize}
\tablewidth{1\textwidth} 
\tablecaption{Stellar Parameters of our sample \label{tab:stellar_parameters}}
\tablehead{
\colhead{Source} & \colhead{2MASS Name} & \colhead{SpT} & \colhead{$T_{\rm eff}$} & \colhead{$L_{\star}$} & \colhead{$M_{\star}$} & \colhead{$\log_{10}(L_{\rm acc})$} & \colhead{$\log_{10}(M_{\rm acc})$} & \colhead{$d$} & \colhead{RUWE} \\
\colhead{} & \colhead{} & \colhead{} & \colhead{(K)} & \colhead{($L_\odot$)} & \colhead{($M_\odot$)} & \colhead{($L_\odot$)} & \colhead{($M_\odot$ yr$^{-1}$)} & \colhead{(pc)} & \colhead{}
}
\colnumbers
\startdata 
J16002612 & J16002612-4153553 & M5.5 & 2920 & 0.08 & 0.11 & -3.2  & -9.6 & 161.5 & 1.103 \\
J16134410 & J16134410-3736462 & M5   & 2980 & 0.04 & 0.10 & -2.5  & -9.0 & 157.1 & 1.152 \\
Sz 73     & J15475693-3514346 & K7   & 4020 & 0.47 & 0.70 & -0.95 & -8.1 & 156.6 & 1.221 \\
Sz 77     & J15514695-3556440 & K7   & 4020 & 0.59 & 0.67 & -1.67 & -8.7 & 154.8 & 1.172 \\
Sz 81A    & J15555030-3801329 & M4.5 & 3085 & 0.24 & 0.19 & -2.4  & -8.9 & 156.8 & 1.314 \\
Sz 81B    & J15555030-3801329 & M5.5 & 3060 & 0.11 & 0.13 & -3.2  & -9.6 & 157.6 & 0.944 \\
Sz 88A    & J16070061-3902194 & M0   & 3900 & 0.31 & 0.65 & -1.4  & -8.5 & 157.2 & 1.204 \\
Sz 97     & J16082180-3904214 & M4   & 3190 & 0.11 & 0.20 & -3.1  & -9.8 & 156.1 & 1.13  \\
Sz 104    & J16083081-3905488 & M5   & 2980 & 0.07 & 0.11 & -3.4  & -9.8 & 158.2 & 1.153 \\
Sz 112    & J16085553-3902339 & M5   & 2980 & 0.12 & 0.14 & -3.4  & -9.8 & 157.7 & 1.091 \\
Sz 130    & J16003103-4143369 & M2   & 3560 & 0.18 & 0.39 & -2.1  & -9.1 & 157.9 & 1.272 \\
\enddata 
\tablecomments{Summary of stellar parameters. Column (1): Source name used in this work. Column (2): 2MASS name. Columns (3)-(8): Spectral type, effective temperature, stellar luminosity, stellar mass, accretion luminosity, and accretion rate \citep{Manara2023}. Column (9): Distance derived from Gaia EDR3 \citep{Bailer-Jones2021}. Column (10): Gaia DR3 RUWE (renormalized unit weight error) \citep{GaiaDR3}.}
\end{deluxetable*}
\onecolumngrid

\begin{deluxetable*}{llllllll}
\tabletypesize{\footnotesize}
\tablewidth{1\textwidth} 
\tablecaption{ALMA Observations Log \label{tab:obs_log}}
\tablehead{
\colhead{Source} & \colhead{Obs. date (start)} & \colhead{Baseline} & \colhead{No. Ant.} & \colhead{Int. time} & \colhead{Phase Cal.} & \colhead{Flux/Bandpass Cal.} & \colhead{PWV} \\
\colhead{} & \colhead{} & \colhead{(m)} & \colhead{} & \colhead{(min)} & \colhead{} & \colhead{} & \colhead{(mm)}
} 
\colnumbers
\startdata 
J16002612   & EB0--3: 05 August 2025 & 175--15238 & 44 & 127.02 & J1610-3958 & EB0--1, 4--5: J1617-5848 & EB0--3: 0.70 \\
            & EB4--5: 06 August 2025 &            &    &        &            & EB2--3: J1427-4206 & EB4--5: 0.55 \\
J16134410   & EB0: 23 July 2025      & 151--15238 & 44 & 62.30  & J1610-3958 & J1427-4206         & EB0: 0.46 \\
            & EB1: 29 July 2025      &            &    &        &            &                    & EB1: 0.60 \\
Sz 73       & EB0: 21 July 2025      & 226--15238 & 42 & 15.42  & J1534-3526 & J1427-4206         & 0.49 \\
Sz 77       & EB0: 23 July 2025      & 233--15238 & 44 & 15.72  & J1534-3526 & J1427-4206         & 0.46 \\
Sz 81A \& B & EB0: 21 July 2025      & 89--12753  & 41 & 15.72  & J1610-3958 & J1427-4206         & 0.60 \\
Sz 88A      & EB0: 29 July 2025      & 151--15238 & 44 & 15.72  & J1610-3958 & EB0: J1924-2914    & EB0: 0.87 \\
            & EB1: 04 August 2025    &            &    &        &            & EB1: J1427-4206    & EB1: 0.53 \\
Sz 97       & EB0: 29 July 2025      & 151--15238 & 44 & 15.72  & J1610-3958 & EB0: J1924-2914    & EB0: 0.87 \\
            & EB1: 04 August 2025    &            &    &        &            & EB1: J1427-4206    & EB1: 0.53 \\
Sz 104      & EB0: 29 July 2025      & 151--15238 & 44 & 15.72  & J1610-3958 & EB0: J1924-2914    & EB0: 0.87 \\
            & EB1: 04 August 2025    &            &    &        &            & EB1: J1427-4206    & EB1: 0.53 \\
Sz 112      & EB0: 29 July 2025      & 151--15238 & 44 & 15.72  & J1610-3958 & EB0: J1924-2914    & EB0: 0.87 \\
            & EB1: 04 August 2025    &            &    &        &            & EB1: J1427-4206    & EB1: 0.53 \\
Sz 130      & EB0: 22 July 2025      & 226--15238 & 45 & 16.03  & J1610-3958 & J1427-4206         & 0.42 \\
\enddata 
\tablecomments{Summary of ALMA observations. Column (1): target name. Column (2): start date of the execution block (EB). Column (3): baseline ranges. Column (4): number of antennas used. Column (5): total on-source integration time. Columns (6) and (7): phase and flux/bandpass calibrators. Column (8): precipitable water vapor (PWV).}
\end{deluxetable*}
\onecolumngrid

\begin{deluxetable*}{lccccccccc}
\tabletypesize{\footnotesize}
\tablewidth{1\textwidth} 
\tablecaption{\galario Best-Fit Results \label{tab:galario_results}}
\tablehead{
\colhead{Source} & \colhead{$\log_{10}(f_0)$} & \colhead{$\sigma_0$} & \colhead{$\log_{10}(f_1)$} & \colhead{$r_1$} & \colhead{$\sigma_1$} & \colhead{inc} & \colhead{PA} & \colhead{$\Delta$RA} & \colhead{$\Delta$Dec}\\
\colhead{} & \colhead{(Jy/sr)} & \colhead{(mas)} & \colhead{(Jy/sr)} & \colhead{(mas)} & \colhead{(mas)} & \colhead{(deg)} & \colhead{(deg)} & \colhead{(mas)} & \colhead{(mas)}
} 
\colnumbers
\startdata 
J16002612 & $10.99^{+0.01}_{-0.01}$ & $9.5^{+0.2}_{-0.2}$ & -- & -- & -- & $19.3^{+4.8}_{-12.5}$ & $132.4^{+25.6}_{-25.7}$ & $0.1^{+0.2}_{-0.2}$ & $-0.7^{+0.2}_{-0.2}$ \\
J16134410 & $10.61^{+0.01}_{-0.01}$ & $18.3^{+0.5}_{-0.5}$ & -- & -- & -- & $32.6^{+3.6}_{-4.1}$ & $155.9^{+7.3}_{-7.2}$ & $0.4^{+0.4}_{-0.4}$ & $-0.1^{+0.4}_{-0.4}$ \\
Sz 73 & $11.434^{+0.003}_{-0.003}$ & $23.5^{+0.1}_{-0.2}$ & $9.22^{+0.02}_{-0.02}$ & $243.4^{+4.8}_{-4.6}$ & $90.9^{+5.5}_{-4.9}$ & $45.2^{+0.5}_{-0.5}$ & $83.8^{+0.6}_{-0.6}$ & $-3.3^{+0.1}_{-0.1}$ & $-3.23^{+0.08}_{-0.08}$ \\
Sz 77 & $11.01^{+0.01}_{-0.01}$ & $20.0^{+0.5}_{-0.4}$ & -- & -- & -- & $50.3^{+1.7}_{-1.8}$ & $105.8^{+2.1}_{-2.1}$ & $1.50^{+0.4}_{-0.4}$ & $-2.5^{+0.3}_{-0.3}$ \\
Sz 81A & $11.16^{+0.1}_{-0.1}$ & $22.7^{+0.2}_{-0.3}$ & -- & -- & -- & $37.9^{+1.1}_{-1.2}$ & $139.9^{+2.2}_{-2.2}$ & $2.2^{+0.2}_{-0.2}$ & $1.6^{+0.2}_{-0.2}$ \\
Sz 81B & $11.07^{+0.02}_{-0.02}$ & $18.2^{+0.5}_{-0.5}$ & -- & -- & -- & $61.7^{+1.7}_{-1.8}$ & $122.8^{+1.8}_{-1.8}$ & $3.1^{+0.4}_{-0.4}$ & $14.7^{+0.4}_{-0.4}$ \\
Sz 88A & $11.458^{+0.007}_{-0.007}$ & $13.9^{+0.2}_{-0.2}$ & -- & -- & -- & $18.80^{+4.2}_{-6.7}$ & $157.1^{+17.5}_{-15.4}$ & $0.09^{+0.2}_{-0.2}$ & $-0.04^{-0.1}_{+0.1}$ \\
Sz 97 & $10.89^{+0.01}_{-0.01}$ & $27.9^{+0.7}_{-0.7}$ & -- & -- & -- & $58.8^{+1.2}_{-1.3}$ & $86.7^{+1.5}_{-1.5}$ & $-2.0^{+0.6}_{-0.6}$ & $1.0^{+0.3}_{-0.3}$ \\
Sz 104 & $11.22^{+0.02}_{-0.02}$ & $11.2^{+0.4}_{-0.3}$ & -- & -- & -- & $22.7^{+4.3}_{-13.0}$ & $24.7^{+28.9}_{-32.7}$ & $2.6^{+0.4}_{-0.3}$ & $-1.3^{+0.3}_{-0.3}$ \\
Sz 112 & -- & -- & $10.91^{+0.02}_{-0.01}$ & $16.2^{+0.4}_{-0.4}$ & $7.7^{+0.4}_{-0.4}$ & $2.91^{+9.92}_{-2.74}$ & $23.3^{+44.1}_{-48.8}$ & $1.8^{+0.4}_{-0.4}$ & $-1.0^{+0.3}_{-0.3}$ \\
Sz 130 & $11.59^{+0.01}_{-0.01}$ & $6.4^{+0.2}_{-0.1}$ & $9.46^{+0.04}_{-0.03}$ & $122.1^{+2.0}_{-1.9}$ & $22.5^{+2.5}_{-2.4}$ & $8.6^{+6.4}_{-5.7}$ & $18.0^{+49.0}_{-46.5}$ & $-0.8^{+0.2}_{-0.2}$ & $1.1^{+0.1}_{-0.1}$ \\
\enddata
\tablecomments{Column (1): target name. Column (2): $\log_{10}$ of flux density $f_0$. Column (3): standard deviation $\sigma_0$. Column (4): $\log_{10}$ of flux density $f_1$. Column (5): radius $r_1$. Column (6): standard deviation $\sigma_1$. Column (7): inclination. Column (8): position angle. Column (9): right ascension offset. Column (10): declination offset. The median of the marginalized posterior distribution is shown, along with the associated statistical uncertainties from the 16th and 84th percentiles of the MCMC marginalized distribution.}
\end{deluxetable*}
\onecolumngrid

\begin{deluxetable}{lccc}
\tabletypesize{\footnotesize}
\tablewidth{1\textwidth}
\tablecaption{Putative planet mass estimates\label{tab:planetmasses} (see Sect.~\ref{subsec:planets})}
\tablehead{
\colhead{Source} & \colhead{$a$} & \colhead{$M_{\rm min}$} & \colhead{$M_{\rm iso}$} \\
\colhead{} & \colhead{(au)} & \colhead{($M_\oplus$)} & \colhead{($M_\oplus$)}
}
\colnumbers
\startdata
Sz 73   & 3.5  & 2.9  & 16.9  \\
Sz 73   & 16.2 & 9.1  & 53.3  \\
Sz 73   & 39.7 & 17.8 & 104.4 \\
Sz 88A  & 2.6  & 2.4  & 13.8  \\
Sz 97   & 3.9  & 4.7  & 25.8  \\
Sz 112  & 0.8  & 1.6  & 8.7   \\
Sz 130  & 8.1  & 6.6  & 37.2  \\
\enddata
\tablecomments{Column (1): target name. Column (2): semi-major axis. Column (3): Minimum gap-opening mass \citep{Sinclair2020}. Column (4): Pebble isolation mass (\citealt{Lambrechts2014}, \citealt{Sinclair2020}).}
\end{deluxetable}
\onecolumngrid

\section{Data reduction and Calibration}
\label{sec:appendix_calibration}

All data were initially calibrated by the ALMA pipeline. Subsequent data treatment was performed in CASA version~6.6.6 \citep{CASATeam2022}. For sources observed across multiple execution blocks (EBs), we verified the astrometric alignment between EBs and found no significant offsets requiring correction, as the phase centers were already consistent to well within the synthesized beam of individual images. No flux rescaling between EBs was needed. No spectral flagging was performed, as no bright lines were expected or detected within our spectral coverage. To reduce data volume, we averaged the visibilities to a channel width of 250\,MHz before concatenating all execution blocks for each target. Since all targets were clearly detected when imaged with robust~2.0 weighting, with signal-to-noise ratios (SNR) between 19 and 78, we performed self-calibration for all sources. We ran one conservative round of phase-only self-calibration with \texttt{gaintype=`T'}, \texttt{combine=`scan,spw'}, and \texttt{solint=`inf'}, yielding SNR improvements of 3--10\%, with an average of 5\%. For the four brightest sources (Sz~73, Sz~81A, Sz~88A, and Sz~130, all with SNR~$>45$), a second round with \texttt{solint=360\,s} provided a further average SNR gain of $\sim 5\%$. For the brightest source in the sample, Sz~73, two additional rounds with \texttt{solint} of 120 and 60\,s were applied, yielding a further improvement of 19\%. After self-calibration, the final measurement sets were time-averaged to 30\,s and frequency-averaged to one channel per spectral window. For the Sz~81 system, we produced two separate measurement sets, one for each component. In each case, the visibilities of the companion were subtracted using the CASA task \texttt{uvsub}, after which the phase center of the Sz~81B measurement set was shifted to the position of the Sz~81B disk.

\section{Dust masses}
\label{sec:appendix_dust_mass}

We report in Table~\ref{tab:image_properties} a lower limit on the total dust mass derived from the integrated flux density via the relation of \citet{Hildebrand1983}, assuming optically thin emission
\begin{equation}
\label{eq:dust_mass}
M_\mathrm{d} = \frac{F_\nu d^2}{B_\nu(T)\, k_\nu},
\end{equation}
where $d$ is the source distance, $B_\nu(T)$ is the Planck function, and $k_\nu = 3.5\,\mathrm{cm^2\,g^{-1}} \times 870\,\mu\mathrm{m}/\lambda$ is the dust opacity \citep{Beckwith1990}. A fixed $T=20$~K is inaccurate for disks as compact as ours, since the midplane temperature $T_\mathrm{mid}(r)$ varies steeply with radius and most of the flux originates within a few au of the star, where the disk is significantly warmer \citep{vanderMarel2023}. We instead adopt the midplane temperature of a passively irradiated disk \citep{Chiang1997}:
\begin{equation}
\label{eq:Tmid}
T_\mathrm{mid}(r) = \left(\frac{\varphi L_\star}{8\pi r^2 \sigma_\mathrm{SB}}\right)^{1/4},
\end{equation}
with $\varphi$ the flaring angle, $L_\star$ the stellar luminosity, and $\sigma_\mathrm{SB}$ the Stefan-Boltzmann constant. We adopt $\varphi=0.02$ \citep{Huang2018_DSHARPII}, and average $T_\mathrm{mid}(r)$ over the observed \frank\ intensity profile $I_\texttt{frank}(r)$ of each disk, truncated at $R_{95}$ (Table~\ref{tab:disk_radii}), so that each radius contributes in proportion to the flux it actually emits:
\begin{equation}
\label{eq:Teff}
\langle T \rangle = \frac{\int_0^{R_{95}} T_\mathrm{mid}(r)\,I_\texttt{frank}(r)\,2\pi r\,\mathrm{d}r}{\int_0^{R_{95}} I_\texttt{frank}(r)\,2\pi r\,\mathrm{d}r}.
\end{equation}
The resulting $\langle T \rangle$ ranges from 30 to 58~K and is used in Eq.~\ref{eq:dust_mass} for the masses reported in Table~\ref{tab:image_properties}. These remain lower limits, as the optically thin assumption is likely to break down in the compact, mostly centrally peaked disks of our sample (see Sect.~\ref{subsec:optical_depth}).

\section{Substructure Identification and Properties}
\label{sec:appendix_subs_prop}

Rings and gaps are defined following the method of \cite{Curone2025}. They are local maxima and minima of the \frank profile, must fall within $R_{90}$, must have a peak intensity above $3\sigma_{\rm rms}$, and are accepted as a gap-ring pair only if the gap depth $I_D/I_B \leq 0.97$, where $I_D$ is the intensity at the gap minimum and $I_B$ is the intensity at the adjacent ring maximum. Substructure widths are derived following \cite{Huang2018_DSHARPII} (see their Section 3.2 and Appendix B for more details): for a gap-ring pair, the dividing point between the outer edge of the gap and the inner edge of the ring, $R_{D,\rm out} \equiv R_{B,\rm in}$, is defined as the radius at which $I(R) = I_{\rm mean} = (I_D + I_B)/2$. The inner edge of the gap, $R_{D,\rm in}$, is the largest radius with $R < R_D$ and $I(R) = I_{\rm mean}$, and the outer edge of the ring, $R_{B,\rm out}$, is the smallest radius with $R > R_B$ and $I(R) = I_{\rm mean}$. The gap and ring widths are then $\Delta R_D = R_{D,\rm out} - R_{D,\rm in}$ and $\Delta R_B = R_{B,\rm out} - R_{B,\rm in}$, respectively. This approach also yields the width of inner cavities: the cavity outer radius corresponds to $R_{B,\rm in}$ of the innermost ring, as for Sz~112. The resulting substructure properties are listed in Table~\ref{tab:substructures}.

Plateaus are identified following \cite{Huang2018_DSHARPII}: contiguous radial segments where $d(\ln I)/dr > -0.2~\rm au^{-1}$ are flagged as plateau candidates. The inner and outer edges of each plateau are labeled D and B at radii $R_D$ and $R_B$, respectively, and each plateau is interpreted as a low-contrast gap-ring pair. The dividing point between the two sub-features, $R_{\rm div} \equiv R_{D,\rm out} = R_{B,\rm in}$, is defined as the radius of maximum $d(\ln I)/dr$ within the plateau, corresponding to the point of flattest slope. The widths of the gap and ring sub-features are $\Delta R_D = R_{\rm div} - R_{D,\rm in}$ and $\Delta R_B = R_{B,\rm out} - R_{\rm div}$, where $R_{D,\rm in}$ and $R_{B,\rm out}$ are defined as the peak of the positive lobe and the trough of the negative lobe of $d^2(\ln I)/dr^2$, respectively. No gap depth is reported for plateau-type features, as no well-defined local minimum exists.

\begin{deluxetable*}{llccccc}
\tabletypesize{\footnotesize}
\tablewidth{1\textwidth}
\tablecaption{Properties of the Continuum Substructures \label{tab:substructures}}
\tablehead{
\colhead{Source} & \colhead{Feature} & \colhead{Radial Location} & \colhead{Width} & \colhead{$R_\mathrm{in}$} & \colhead{$R_\mathrm{out}$} & \colhead{Depth} \\
\colhead{} & \colhead{} & \colhead{(au)} & \colhead{(au)} & \colhead{(au)} & \colhead{(au)} & \colhead{}
}
\colnumbers
\startdata
\multicolumn{7}{l}{\textit{Inner disk substructures}} \\[2pt]
Sz 73 & D4 & $3.5$ & $1.3$ & $3.3$ & $4.6$ & -- \\
      & B6 & $5.5$ & $2.2$ & $4.6$ & $6.8$ & -- \\
\noalign{\vskip 4pt}
Sz 88A    & D3 & $2.6$ & $1.0$ & $2.2$ & $3.2$ & -- \\
          & B4 & $3.7$ & $1.2$ & $3.2$ & $4.4$ & -- \\
\noalign{\vskip 4pt}
Sz 97     & D4 & $ 3.9$ & $1.4$& $ 3.0$ & $4.4$ & $0.85$ \\
          & B6 & $5.7$ & $1.8$ & $4.4$ & $6.4$ & -- \\
\noalign{\vskip 4pt}
Sz 112    & B3 & $2.6$ & $2.0$ & $1.6$ & $3.6$ & -- \\[4pt]
\hline
\noalign{\vskip 4pt}
\multicolumn{7}{l}{\textit{Outer disk substructures}} \\[2pt]
Sz 73     & D16 & $16.2$ & $19.4$ & $10.0$ & $29.4$ & $0.13$ \\
          & B33 & $33.1$ & $6.0$ & $29.4$ & $35.4$ & -- \\
          & D40 & $39.7$ & $7.9$ & $36.2$ & $44.1$ & $0.16$ \\
          & B47 & $47.3$ & $5.8$ & $44.1$ & $49.9$ & -- \\
\noalign{\vskip 4pt}
Sz 130    & D8  & $8.1$ & $11.6$ & $3.6$ & $15.2$ & $0.0003$ \\
          & B18 & $18.0$ & $8.2$ & $15.2$ & $23.4$ & -- \\
\enddata
\tablecomments{Column (1): target name. Column (2): annular substructure label. "B" (for bright) indicates a ring, while "D" (for dark) indicates a gap. The number in the label is the feature distance from the central star measured in au. Column (3): substructure radial location. Column (4): annular substructure width. Columns (5) and (6): inner and outer edge of the substructure. Column (7): gap depth.}
\end{deluxetable*}
\onecolumngrid

\section{Comparison with Bae et al. (2023)}
\label{sec:comparison_bae2023}

\begin{figure}
\centering
\includegraphics[width=0.5\hsize]{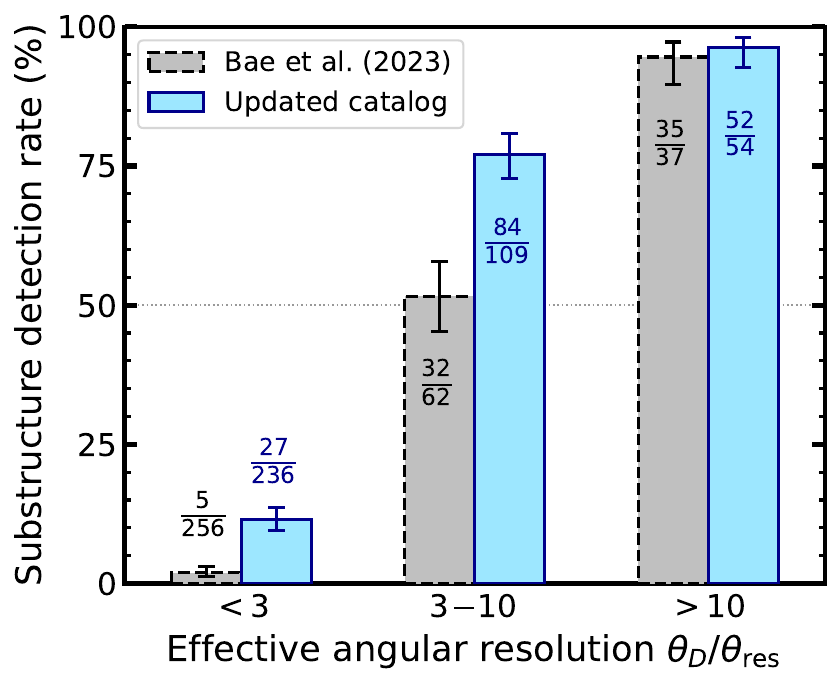}
\caption{Substructure detection rate as a function of effective angular resolution, $\theta_D/\theta_{\rm res}$, comparing the sample of \citet{Bae_2023_PPVII} with our updated catalog. The bins follow those defined in \citet{Bae_2023_PPVII}. Uncertainties represent $1\sigma$ Wilson score confidence intervals. Each bar is labeled with the corresponding ratio of disks with detected substructure to the total in that bin.}
\label{fig:comparison_Bae}
\end{figure}

Figure~\ref{fig:comparison_Bae} compares the substructure detection rate as a function of $\theta_D/\theta_{\rm res}$ between \citet{Bae_2023_PPVII} and our updated catalog, using the same bins defined in \citet{Bae_2023_PPVII}. The updated catalog contains a larger total number of disks (399 vs.\ 355), owing to new ALMA observations added since \citet{Bae_2023_PPVII}, and a higher fraction of sources with $\theta_D/\theta_{\rm res} > 3$, reflecting the increasing number of high-resolution surveys. The detection rate in the highest effective resolution bin ($\theta_D/\theta_{\rm res} > 10$) is unchanged, while it increases noticeably in the two lower effective resolution bins. This increase is driven in large part by more recent works that classified disk substructure not solely from visual inspection of CLEAN images, but also from visibility analysis and super-resolution imaging techniques, which are more sensitive to low-contrast features.

\bibliography{bibliography}{}
\bibliographystyle{aasjournalv7}



\end{document}